\documentclass[preprint2,twocolumn,times,tighten]{aastex63}

\usepackage{graphicx,times}
\usepackage{subfigure}

\usepackage{threeparttable}
\usepackage{booktabs}

\usepackage{amsmath}
\usepackage{cases}
\usepackage{longtable}
\usepackage{hyperref}
\usepackage{epstopdf}
\usepackage{amsmath,bm}
\usepackage{amssymb}
\usepackage{natbib}
\usepackage{morefloats}
\usepackage{multirow}
\usepackage{array}
\usepackage{verbatim}
\usepackage{color}
\usepackage{lineno}
\begin{document}

\title{Turbulent reconnection acceleration}


\author[0000-0002-0458-7828]{Siyao Xu}
\affiliation{Institute for Advanced Study, 1 Einstein Drive, Princeton, NJ 08540, USA; sxu@ias.edu
\footnote{NASA Hubble Fellow}}


\author{Alex Lazarian}
\affiliation{Department of Astronomy, University of Wisconsin, 475 North Charter Street, Madison, WI 53706, USA; 
lazarian@astro.wisc.edu}

\affiliation{Centro de Investigación en Astronomía, Universidad Bernardo O’Higgins, Santiago, General Gana 1760, 8370993,Chile}

\begin{abstract}
The ubiquitous turbulence in astrophysical plasmas is important for both magnetic reconnection and reconnection acceleration. 
We study the particle acceleration during 
fast 3D turbulent reconnection with reconnection-driven turbulence. 
Particles bounce back and forth between the reconnection-driven inflows 
due to the mirror reflection and convergence of strong magnetic fields.
Via successive head-on collisions, the kinetic energy of the inflows is converted into 
the accelerated particles. 
Turbulence not only regulates the inflow speed 
but also introduces various inflow obliquities with respect to the local turbulent magnetic fields. 
As both the energy gain and escape probability of particles 
depend on the inflow speed, the spectral index of particle energy spectrum is not universal.
We find it in the range from $\approx 2.5$ to $4$, with the steepest spectrum expected at a strong guide field, i.e. a small angle between the total incoming magnetic field and the guide field. 
Without scattering diffusion needed for confining particles, 
the reconnection acceleration can be very efficient at a large inflow speed and a weak guide field. 

\end{abstract}


\section{Introduction}

Reconnection acceleration is a fundamental acceleration process in solar and space plasmas, as well as in high-energy 
astrophysical systems 
\citep{LucS18,Mar20,Laz20,Shib20,Li21}.
It accounts for the particle energization in diverse environments, such as 
the solar atmosphere
\citep{TUrk05,Chi20,Jaf21},
the Heliosheath
\citep{LaO09}, 
pulsar wind nebulae 
\citep{Lyut18}, 
gamma-ray bursts 
\citep{Zh11,Deng16,LazZ19}, 
accretion disks and jets
\citep{DeG05,Kado21}.

As demonstrated by both magnetohydrodynamic (MHD) simulations and 
particle-in-cell (PIC) simulations, reconnection and reconnection acceleration in 3D space are intrinsically different from their 2D models 
\citep{Prie03,Kow11,Parn11,Dah17,Wern21,Zhg21,Zhang21}.
As a result of reduction in dimensionality, 
reconnection in 2D occurs at single X-type magnetic null points.
Reconnection acceleration in 2D has limited efficiency and energy gain 
due to the insufficient global volume-averaged reconnection rate   
and the trapping of particles within 2D magnetic islands.
In realistic 3D astrophysical plasmas, 
magnetic fields are not confined in a closed domain, and
the ubiquitous turbulence regulates the diffusion of magnetic fields.
Consequently, the reconnection occurs continuously throughout a volume determined by the extent of the diffusion of turbulent magnetic fields
(\citealt{LV99}; see
\citealt{Laz20} for a review). 
Turbulence makes magnetic field lines intrinsically stochastic \citep{Eyin13}, introducing stochasticity in  particle propagation and new diffusion mechanism of particles 
\citep{LX21m}.


The 3D turbulent reconnection model proposed by 
\citet{LV99} (henceforth LV99)
describes a fast reconnection process in the sense that the 
reconnection rate does not depend on the microscopic resistivity
\citep{Eyink2011}.
It is generally believed that the macroscopic transport properties of turbulence are independent of microscopic diffusivities
\citep{Spie71}.
This is also found to be true for diffusion of magnetic fields in the presence of turbulence 
\citep{LVC04,Sant10,Laz14r}.
Efficient reconnection acceleration requires a sufficiently high reconnection rate.
Although a localized high reconnection rate can be achieved by adding new microscopic physics at 
an X-point
(e.g., \citealt{Par93,Man94,Dra08}), 
it is not clear how to explain a high global reconnection rate on macroscopic scales required for efficient acceleration of energetic particles.
With turbulence-governed diffusion of magnetic fields, 
the LV99 model naturally has a high global reconnection rate when the level of turbulence is high. 
The turbulence can be driven by diverse astrophysical processes, 
various instabilities, or reconnection itself 
\citep{Uri17,Kow17,Kow20}.
In the case of reconnection-driven turbulence, 
the level of turbulence depends on the fluid compressibility 
\citep{Kow17}, 
and a reconnection rate on the system size comparable to $V_A$ can be reached 
(Kowal et al. in prep), 
where $V_A$ is the Alfv\'{e}n speed.

On the basis of the LV99 model of turbulent reconnection, 
\citet{DeG05}
proposed that particles experience successive head-on collisions with the inflows when they are trapped by converging magnetic mirrors 
during turbulent reconnection. 
Efficient particle acceleration in simulations of turbulent reconnection was found by 
\citet{Kow11,Kow12,Zjf22}.
\citet{Bere16}
considered the curvature drift acceleration during 
the initial stage of spontaneous turbulent reconnection with { insignificant turbulence and}
negligible inflows. 
For the particle acceleration during turbulent reconnection with a global high reconnection rate and inflow speed, 
a quantitative description is still missing. 
Moreover, because of the limited ranges of length scales and particle energies, and the partially developed turbulence in PIC simulations, 
the effect of turbulence on the 
simulated reconnection and reconnection acceleration cannot be fully investigated, and the energy gain of particles is also limited, 
e.g., up to $100$ MeV for electrons even for relativistic reconnection
\citep{Sir14,Guo14}.
These are the motivations of the current theoretical study on turbulent reconnection acceleration.

The reconnection acceleration of particles with the gyroradii larger than the thickness of the reconnection region was 
theoretically proposed by 
\citet{oiLa12}
and numerically demonstrated by 
\citet{Zhang21}. 
Here we focus on the case with the gyroradii smaller than the thickness of the reconnection region
{ and the acceleration process inside the reconnection region}. 
In Section 2, we first briefly review the basic physical pictures of turbulent reconnection and reconnection acceleration. 
In Section 3, we study in detail the acceleration of both reflected and unreflected particles, and their averaged fractional energy gain. 
In Sections 4 and 5, we analyze the escape probability and energy spectral index of reconnection accelerated particles, as well as 
the acceleration timescale. 
Discussion and final conclusions are provided in Sections 6 and 7.

\section{Turbulent reconnection and reconnection acceleration}

\subsection{Turbulent reconnection of magnetic fields }
\label{ssec:turrec}

The reconnection of non-turbulent magnetic fields happens due to the resistive diffusion of magnetic fields, and thus the 
reconnection rate is limited by the resistivity 
\citep{Pars57,Swe58}, with 
\begin{equation}
    \frac{U_\text{in}}{V_A} = S^{-\frac{1}{2}} = \frac{\Delta}{L_x}, 
\end{equation}
where 
$U_\text{in}$ is the inflow speed that characterizes the speed of reconnection, 
\begin{equation}
   V_A = U_\text{out} = \frac{B_\text{in} \sin\theta}{\sqrt{4\pi\rho}}
\end{equation}
is the Alfv\'{e}n speed (equal to the outflow speed) corresponding to the strength $B_\text{in}\sin\theta$
of oppositely directed magnetic fields in the inflows, 
$B_\text{in}$ is the total magnetic field strength in the inflows, 
$\theta$ is the angle between ${\bm B_\text{in}}$ and the guide field ${\bm B_\text{g}}$, 
the guide field strength is $B_g = B_\text{in} \cos \theta$, 
$\rho$ is the plasma density that is constant in incompressible medium considered in this work, 
\begin{equation}
   S = \frac{L_x V_A}{\eta}
\end{equation}
is the Lundquist number, 
$L_x$ and $\Delta$ are the length and thickness of the reconnection region, 
and $\eta$ is the resistive diffusivity.
Because of the slow resistive diffusion and thus large disparity between $L_x$ and $\Delta$, 
the Sweet-Parker (SP) model of reconnection is slow with $U_\text{in} \ll V_A$.

\begin{figure*}[ht]
\centering
\subfigure[]{
   \includegraphics[width=8cm]{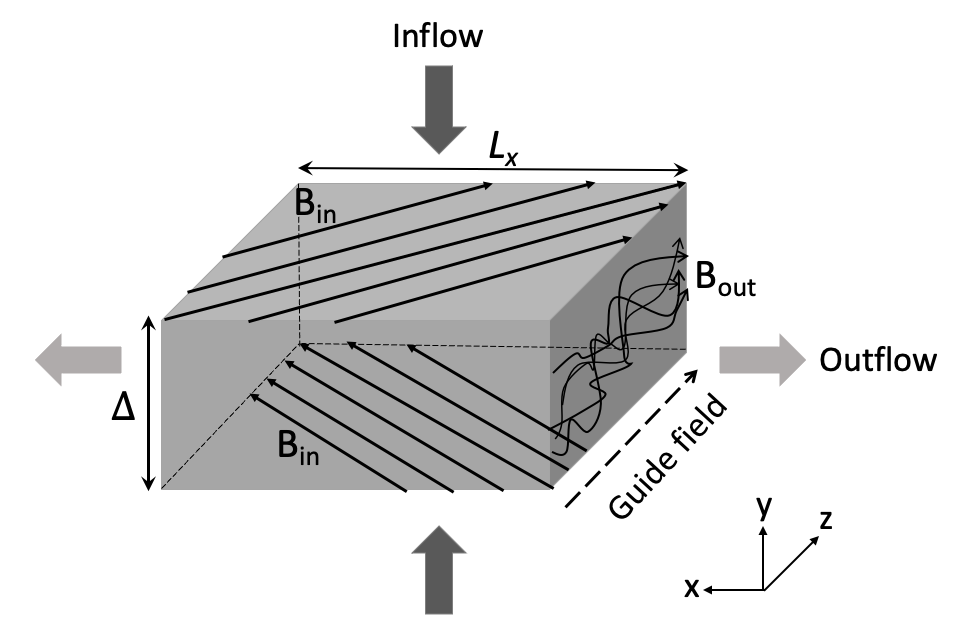}\label{fig: 3d}}
\subfigure[]{
   \includegraphics[width=7cm]{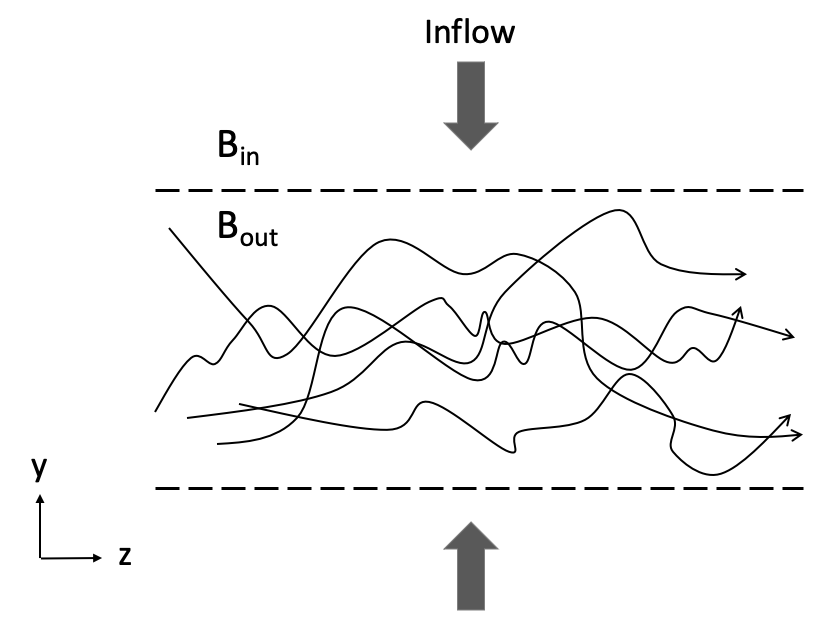}\label{fig: yv}}
\caption{Illustration for turbulent magnetic reconnection with reconnection-driven turbulence. The dashed arrow in 
(a) indicates the initial guide field.
Via reconnection, the guide field becomes turbulent as seen in the $y-z$ plane in (b). 
The region between the dashed lines in (b) is the reconnection region.}
\label{fig: rec}
\end{figure*}

For turbulent magnetic fields, their diffusion is regulated by turbulence rather than resistivity
(LV99; \citealt{Eyink2011}). 
Therefore, $\Delta$ is determined by the diffusion distance of turbulent magnetic fields in $y$ direction
corresponding to a distance $L_x$ in $x$ direction
(see Fig. \ref{fig: rec}).
The resulting reconnection rate is 
\citep{LV99}
\begin{equation}\label{eq: lvrer}
    \frac{U_\text{in}}{V_A} = \text{min} \bigg[\Big(\frac{L_x}{L_i}\Big)^\frac{1}{2}, \Big(\frac{L_i}{L_x}\Big)^\frac{1}{2}\bigg] M_A^2, ~~M_A<1,
\end{equation}
where $L_i$ is the correlation length of the externally driven turbulence that can be larger or smaller than $L_x$, 
and $M_A = V_L/V_A$ is defined as the Alfv\'{e}n Mach number with the turbulent velocity $V_L$ at $L_i$ and $V_A$ 
corresponding to $B_\text{in}\sin\theta$. 
When $L_i\sim L_x$ and $M_A \sim 1$, the LV99 model of turbulent reconnection leads to $U_\text{in} \sim V_A$.
When $L_i$ and $L_x$ are very different or $M_A\ll1$ for weakly turbulent magnetic fields, 
there can be $U_\text{in}\ll V_A$. 
The LV99 model describes a fast reconnection process in the sense that the reconnection rate is independent of the microscopic magnetic diffusivity. 
This is simply because the process of turbulent (reconnection) diffusion of magnetic fields, frequently referred to as reconnection diffusion
\citep{Laz05},
is faster than their microscopic diffusion on scales where turbulence exists. 
The rigorous mathematical proof of the above statement is provided in
\citet{Eyink2011,Eyink15}.
The predicted fast reconnection 
has been confirmed by 3D MHD turbulent reconnection simulations in both non-relativistic
\citep{KowL09,KL12} and  relativistic
\citep{Tak15} 
cases.

In addition to the ubiquitous turbulence in astrophysical plasmas, 
reconnection itself naturally drives turbulence with the released magnetic energy converted to turbulent energy. 
The reconnection-driven turbulence in turn regulates the reconnection rate 
(LV99; \citealt{LV00}).
The reconnection process with reconnection-driven turbulence is illustrated in Fig. \ref{fig: rec}. 
In the steady state, the antiparallel component of ${\bm B_\text{in}}$ in the inflows reconnect. 
The reconnection-driven turbulence determines $U_\text{in}$ and the outflow thickness $\Delta$, 
and causes perturbation of the guide field that is initially along the $z$ direction. 
The resulting turbulent guide field ${\bm B_\text{out}}$ in the outflow leaves the system along the $x$ direction
with the outflow speed $U_\text{out} = V_A$. 
This process has been numerically demonstrated by 
\citet{Kow17,Kow20}
with 3D MHD simulations.
The turbulence generated during reconnection can originate from the Kelvin-Helmholtz instability induced by velocity shear in 
reconnecting layers, 
which dominates over the tearing instability in driving turbulence 
\citep{Kow20}.

For the magnetic reconnetion in a compressible medium, 
the conservation of mass leads to 
\begin{equation}
    \frac{U_\text{in}}{V_A} = \frac{\rho_\text{out}}{\rho_\text{in}} \frac{\Delta}{ L_x}, 
\end{equation}
where $\rho_\text{in}$ and $\rho_\text{out}$ are the densities in the inflows and outflows, respectively. 
Given $\rho_\text{out}>\rho_\text{in}$, 
a higher reconnection rate in a compressible medium is expected compared to that in an incompressible medium. 
The density change and density inhomogeneity in a compressible medium 
can enhance the growth of instabilities, e.g., 
the Rayleigh-Taylor instability, the Richtmyer-Meshkov instability,
that do not exist in an incompressible medium and thus lead to a higher 
level of turbulence. 
\citet{Kow17}
found that in a compressible medium, 
the reconnection-driven turbulence and the resulting $U_\text{in}$ have dependence on plasma $\beta_p (= \text{gas pressure}/\text{magnetic pressure})$ given the same free magnetic energy. 


In our analysis, we consider  
\begin{equation}\label{eq: velin}
   U_\text{in} = f V_A = fU_\text{out}= f \frac{B_\text{in} \sin \theta}{\sqrt{4\pi\rho}} 
\end{equation}
for turbulent magnetic reconnection,
where $\theta$ is the angle between the initial guide field and ${\bm B_\text{in}}$.
The factor $f = \Delta / L_x \lesssim1$ depends on turbulent properties (see Eq. \eqref{eq: lvrer}) and plasma $\beta_p$
in the case of a compressible medium. 
$f$ in the range $0.001-0.2$ is indicated by observations of solar flares 
\citep{Cass17}.
While a small $f$ up to $0.02$ is seen in 
weakly compressible reconnection simulations with reconnection-driven turbulence 
\citep{Kow17},
a large $f$ with values above $0.1$ is found in a highly compressible medium 
(Kowal et al. in prep).


For simplicity, we consider that the fluctuation component of $B_\text{out}$ comes from the reconnection-driven turbulence alone,
with 
\begin{equation}
   \delta B_\text{out} \sim B_\text{in} \sin\theta.
\end{equation}
We define the local obliquity $\alpha$ of the inflow 
as the angle between the inflow direction ($y$ direction) and the local ${\bm B_\text{out}}$ direction.
The original inflow obliquity with respect to the initial guide field is $\sim 90^\circ$ (see Fig. \ref{fig: rec}(b)).
Due to the reconnection-driven turbulence, 
the inflow has a range of local obliquities $(90^\circ- \delta \alpha, 90^\circ)$. 
For the variation of obliquity $\delta \alpha$, there is  
\begin{equation}
   \tan \delta \alpha = \frac{\delta B_\text{out}}{B_g} \sim \frac{B_\text{in}\sin\theta}{ B_\text{in} \cos\theta}
   = \tan \theta,
\end{equation}
where $B_g$ is the strength of initial guide field. 
Therefore, we approximately have $\delta \alpha \sim \theta$.


\subsection{Reconnection acceleration of particles}
\label{ssec:recacc}

Reconnection acceleration of particles happens due to the convergence of flows driven by the reconnection, 
irrespective of the compressibility of the medium. 
As the particles are confined in the reconnection region by mirror reflection or 
gyrating around the converging magnetic fields, 
unlike the diffusive shock acceleration (DSA, 
\citealt{Axf77,Kry77,Bell78,Blan78,Drury83,Jon91}), 
scattering diffusion is not a necessary confining mechanism 
for reconnection acceleration. 
Efficient reconnection acceleration requires a sufficiently high reconnection rate and thus a sufficiently high level of turbulence. 
The associated electric field for the acceleration is induced by the inflow motion of plasma, while the 
effect of Ohmic/non-ideal electric field in microscopic diffusion regions is negligible
\citep{Kow12,Guo19}.

Although there is a divergence of flow associated with the outflows from the reconnection region, 
it does not necessarily cause energy loss of particles.
For particles with the gyroradii $r_g$ smaller than $\Delta$, 
they cannot freely diffuse in the $x$ direction (see Fig. \ref{fig: rec}) 
by crossing the guide field (along $z$) and the transverse magnetic field (along $y$) in the reconnection region. 
\citet{Druf12}
considered the case with $r_g > \Delta$. 
One of the main assumptions in 
\citet{Druf12}
is that the particles with isotropic distribution can sample the entire system that encompasses both inflows and outflows. 
Under this assumption, compression of plasma becomes a necessary condition for acceleration to happen. 
However, as we mentioned, this assumption is not valid for particles with $r_g < \Delta$. 
For particles with $r_g > \Delta$, to satisfy this assumption, 
very efficient scattering is required on scales larger than $\Delta$ to ensure the isotropic distribution and free diffusion of particles.
{ As reconnection-driven turbulence only exists within the reconnection region on scales $\lesssim \Delta$, }
the origin of the scattering is unclear.

For particles with $r_g < \Delta$, 
they bounce back and forth within the reconnection region 
and undergo the first-order Fermi acceleration by colliding with the inflows many times 
\citep{DeG05}.
The particles entrained on the magnetic fields in the outflow are advected away from the reconnection region 
in either direction along the $x$-axis (Fig. \ref{fig: rec}).
As the particles cannot freely diffuse and simultaneously sample the bidirectional outflows, 
they do not suffer the energy loss due to the flow divergence. 
Efficient particle acceleration during turbulent reconnection has been numerically demonstrated in 
\citet{Kow11,Kow12,Zjf22}.

The curvature drift acceleration is the main acceleration mechanism in 2D reconnection models. 
The acceleration happens due to the curvature
drift of particles in the direction of the motional electric field, which is induced by island contraction or merging.
Particles are trapped within a contracting magnetic island or between two merging islands
\citep{Drak06,Drak13,leRZ18}.
The acceleration stops when contraction or merging comes to an end. 
In 3D turbulent reconnection,
{ turbulent magnetic fields do not 
form 2D islands.
The curvature of magnetic fields caused by turbulence results in stochastic acceleration with both}
curvature drift acceleration and deceleration 
\citep{Bere16}.
{ The release of magnetic energy via relaxation of magnetic tension of curved field lines is found to be subdominant compared to that via magnetic pressure gradient when the 3D reconnection evolves to a more turbulent state
\citep{Du22}.
Given the expected stochastic nature of the curvature drift acceleration and subdominant magnetic energy release via curvature relaxation
in turbulent reconnection,} 
we 
focus below on the systematic acceleration associated with { the magnetic field gradient} in inflows.

\begin{figure}[ht]
\centering
   \includegraphics[width=7cm]{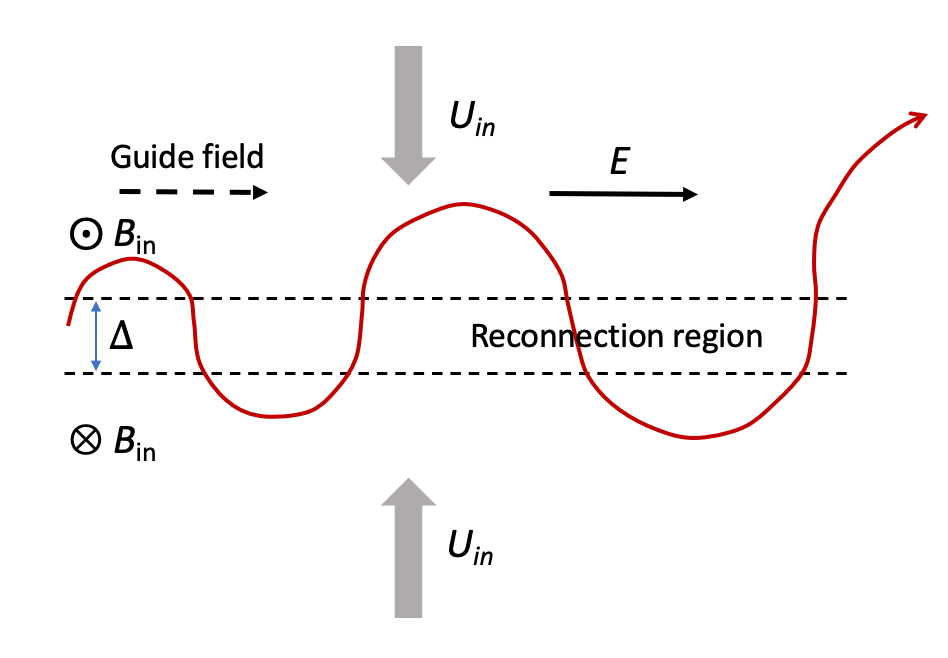}
\caption{Illustration for reconnection acceleration with  
$r_g > \Delta$. ${\bm E} = - {\bm U_\text{in}} \times {\bm B_\text{in}}/c$ is the electric field calculated in the rest frame of the plasma
in the reconnection region.
The drift of particle is caused by the different magnetic field directions in the inflows above and below the reconnection region.
The drift direction coincides with the initial guide field direction and ${\bm E}$ direction, and thus the particle is accelerated 
during every gyration. 
The original picture was presented in \citet{oiLa12}.}
\label{fig: gyrol}
\end{figure}

For particles with $r_g > \Delta$, they gyrate around the reconnection region. 
As illustrated in Fig. \ref{fig: gyrol},
because of the opposite directions of reconnecting magnetic fields, 
particles drift along the $z$ direction, parallel to the motional electric field 
${\bm E} = - {\bm U_\text{in}} \times {\bm B_\text{in}}/c$, where $c$ is the light speed. 
This mechanism was first proposed in 
\citet{oiLa12}
and recently confirmed by 
\citet{Zhang21}
with 3D PIC simulations of relativistic reconnection.

In this work,
we will focus on the particle acceleration with $r_g < \Delta$ during turbulent reconnection.

\section{Particle energy gain}

\subsection{Reflected particles }
\subsubsection{Energy gain per reflection}
\label{sssec: enerefl}

\begin{figure}[ht]
\centering
   \includegraphics[width=9cm]{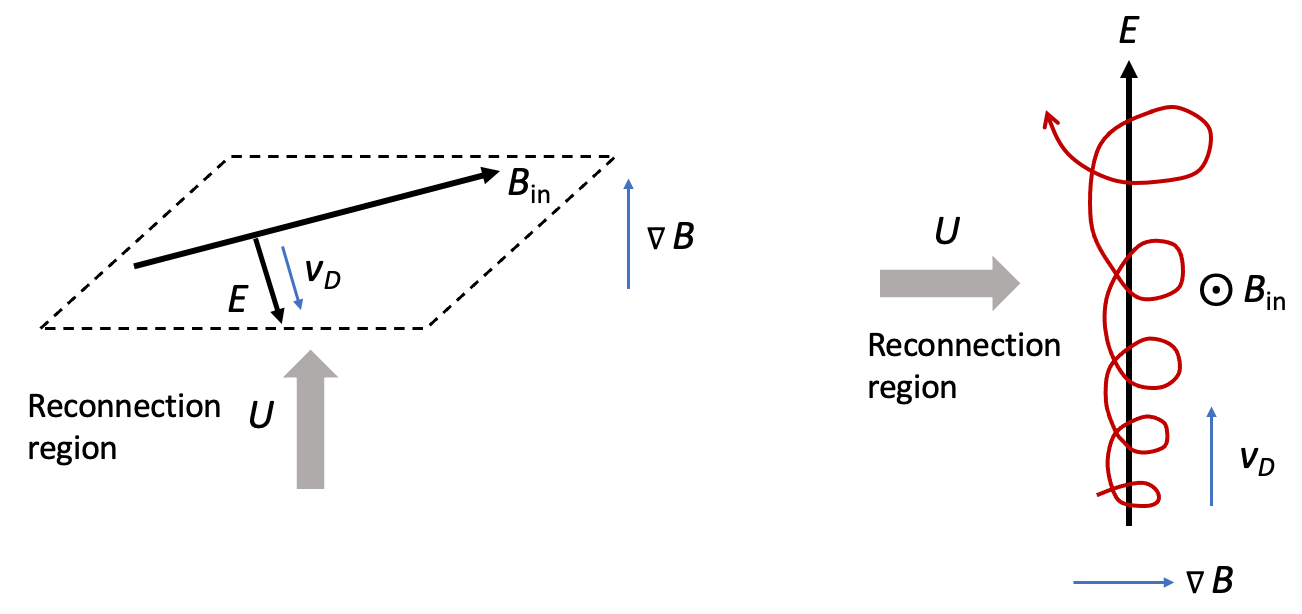}
\caption{Illustration for the RDA with  
$r_g < \Delta$. Left and right panels show one encounter of a particle with the inflow seen from different viewing angles.
$U (=U_\text{in})$ is the plasma speed in the reconnection region seen in the reference frame of the inflow. 
${\bm E} = - {\bm U} \times {\bm B_\text{in}}/c$ is the motional electric field. 
The magnetic field gradient $\nabla B$ is caused by the transition from the strong magnetic field in the inflow to the weak magnetic field in the reconnection region. The $\nabla B$ drift of a particle with the drift speed $v_D$ is along the ${\bm E}$ direction.
Its trajectory is shown in the right panel (red).
A particle in the reconnection region experiences multiple encounters with the inflows and thus is repeatedly accelerated. 
}
\label{fig: gyros}
\end{figure}

As mentioned above, 
with respect to the turbulent guide field ${\bm B_\text{out}}$, 
the inflow velocity along the local magnetic field direction becomes   
$U_\text{in}/\cos\alpha$,
where $U_\text{in}$ is the inflow speed along the $y$ direction. 
Similar to the shock drift acceleration (SDA) at an oblique shock
\citep{Son69,Wu84,Arm85,BaMel01,XL22}, 
because of the change of magnetic field strength from $B_\text{out}$ in the reconnection region to $B_\text{in}$ in the inflow, 
a particle incident on the inflow experiences a gradient-B drift along the induced electric field 
$-{\bm U} \times {\bm B_\text{in}}/c$ in the inflow frame, where ${\bm U}$ is the flow velocity seen in the inflow frame. 
We term this acceleration process ``reconnection drift acceleration (RDA)" (see Fig. \ref{fig: gyros}).
During the RDA, the kinetic energy of the inflow is converted to the particle kinetic energy.

{ For shock acceleration, the SDA, i.e., the gradient-B drift acceleration, is equivalent to the acceleration via reflection by approaching mirrors, i.e., the Fermi mechanism 
\citep{Lee82,Krauss89}.
As the shock front, i.e., the magnetic field gradient/magnetic mirror is not at rest, it is convenient to adopt a special reference frame for which 
the intersection point of the shock front and the magnetic field line is at rest, i.e., 
the de-Hoffmann-Teller (HT) frame
\citep{deTe50}. 
The fractional energy gain can then be obtained by transformation back to the rest frame of the upstream medium
\citep{Ostr88,XL22}. }

{ Similarly, 
for reconnection acceleration, 
the RDA, i.e., the gradient-B drift acceleration, is equivalent to the acceleration via reflection by approaching magnetic mirrors.
Despite the similarity in the acceleration mechanism, the magnetic mirror does not form due to 
shock compression, but the annihilation of oppositively directed magnetic fields. It creates the 
magnetic pressure gradient that drives inflows and approaching magnetic mirrors. }
To derive the energy gain during the interaction with the inflow, 
we also adopt the 
HT frame as long as the inflow is subluminal.
In the HT frame, 
the intersection point between the inflow and the local guide field is at rest 
and the fluid motion is parallel to the magnetic field. 
Therefore, the electric field vanishes, and the particle energy remains unchanged in the HT frame. 
Given the similarity between RDA and SDA, 
here we follow a similar analysis to that in \citet{Ostr88} for SDA. 
The Lorentz factor $\gamma$ of a particle in the rest frame of the reconnection region
is related to $\gamma^\prime$ in the HT frame by 
\begin{equation}\label{eq: gamgamp}
    \gamma = \Gamma \gamma^\prime (1 - \beta \mu^\prime), 
\end{equation}
where $\Gamma = 1/\sqrt{1-\beta^2}$, and 
\begin{equation}\label{eq: expbet}
   \beta = \frac{U_\text{in}}{c\cos \alpha }.
\end{equation}
The cosine of the particle pitch angle $\mu$ in the rest frame of the reconnection region
is related to $\mu^\prime$ in the HT frame by 
\begin{equation}\label{eq:mumup}
     \mu = \frac{\mu^\prime - \beta}{1 - \beta \mu^\prime}.
\end{equation}
A particle incident from the reconnection region on the inflow can be 
reflected by the mirror force due to the increase of magnetic field strength from the reconnection region to the inflow. 
Under the consideration of conserved magnetic moment in the HT frame, the relation 
\begin{equation}
    \frac{{v^\prime}_\perp^2}{B_\text{out}} = \frac{{v^\prime}^2}{B_\text{in}}
\end{equation}
constrains the smallest pitch angle in the HT frame for reflection to happen,
where ${v^\prime}_\perp$ is the particle velocity perpendicular to the magnetic field in the HT frame, and 
$v^\prime$ is the total particle velocity in the HT frame.
We find that particles with $\mu^\prime$ in the range 
\begin{equation}
   0 < \mu^\prime < \sqrt{1 - \frac{B_\text{out}}{B_\text{in}}}
\end{equation}
can be reflected. It corresponds to (Eq. \eqref{eq:mumup})
\begin{equation}\label{eq: rmofr}
    -\beta < \mu < \frac{\sqrt{1-\frac{B_\text{out}}{B_\text{in}}}-\beta}{1 - \beta \sqrt{1- \frac{B_\text{out}}{B_\text{in}}}}
\end{equation}
for the range of $\mu$.
Compared to the mirroring effect due to shock compression of magnetic fields, 
at a large $\theta$ with a weak guide field, 
the mirroring effect caused by magnetic reconnection is much stronger, and thus particles with a large range of $\mu$ can be reflected. 
For the unreflected particles in the loss cone with 
\begin{equation}\label{eq:loscmur}
        \frac{\sqrt{1-\frac{B_\text{out}}{B_\text{in}}}-\beta}{1 - \beta \sqrt{1- \frac{B_\text{out}}{B_\text{in}}}} <\mu < 1,
\end{equation}
unlike the case of an oblique shock, 
instead of being transmitted, they are attached to the incoming magnetic fields via gyromotion and further advected back to the 
reconnection region together with the converging magnetic fields carried by the inflow.
Both reflected and unreflected particles undergo the RDA.
Here we will first analyze the acceleration of reflected particles and move to the acceleration 
of unreflected particles in Section \ref{ssec: unrefl}.

Given the symmetry of the inflow, when 
\begin{equation}
   \frac{\sqrt{1-\frac{B_\text{out}}{B_\text{in}}}-\beta}{1 - \beta \sqrt{1- \frac{B_\text{out}}{B_\text{in}}}}>0,
\end{equation}
that is 
\begin{equation}\label{eq: conrfleb}
   \beta < \sqrt{1-\frac{B_\text{out}}{B_\text{in}}},
\end{equation}
we consider the range of $\mu$
\begin{equation}\label{eq: rmofrsym}
    0 < \mu < \frac{\sqrt{1-\frac{B_\text{out}}{B_\text{in}}}-\beta}{1 - \beta \sqrt{1- \frac{B_\text{out}}{B_\text{in}}}}
\end{equation}
for reflection to happen. 
For a reflected particle, the ratio of the energies after ($E_f$) and before ($E_0$) the reflection is 
(Eq. \eqref{eq: gamgamp})
\begin{equation}\label{eq: efe0}
   \Big(\frac{E_f}{E_0}\Big)_\text{r} = 
   \frac{1+ \beta \mu^\prime}{1-\beta \mu^\prime},
\end{equation}
where $\mu^\prime$ and $-\mu^\prime$ are the pitch angle cosines before and after the reflection in the HT frame. 
By using the relation in Eq. \eqref{eq:mumup}, Eq. \eqref{eq: efe0} leads to 
\begin{equation}\label{eq: enf0}
    \Big(\frac{E_f}{E_0} \Big)_\text{r}
    = \frac{1+2\beta \mu + \beta^2}{1-\beta^2}.
\end{equation}
In the limit of a small $\beta$, 
the above expression approximately becomes 
\begin{equation}
    \Big(\frac{E_f}{E_0}\Big)_\text{r} \approx 1 + 2 \beta \mu .
\end{equation} 
At a large $\beta$, the quadratic term in Eq. \eqref{eq: enf0} is not negligible, 
and the energy gain after one reflection can be very significant. 
The acceleration of reflected particles can happen over a range of pitch angles and inflow obliquities. 
To obtain the particle energy spectrum, we will need both pitch-angle and obliquity averaged fractional energy gain of reflected particles 
(see Section \ref{ssec: averefunr}).

\subsubsection{Pitch angle distribution}
\label{ssec:ani}

As a particle reverses its direction in the HT frame after reflection, by using 
Eq. \eqref{eq:mumup}, we can obtain 
the pitch angle cosine after reflection 
\begin{equation}\label{eq: mufinc}
    \mu_f =\frac{-\mu^\prime- \beta}{1+\beta \mu^\prime}
    = -\frac{\mu+2\beta + \beta^2 \mu}{ 1+2\beta \mu + \beta^2}.
\end{equation}
We find that the ratio of the particle momentum perpendicular to the magnetic field after and before the reflection as 
(Eqs. \eqref{eq: enf0} and \eqref{eq: mufinc})
\begin{equation}
    \frac{p_{\perp,f}}{p_{\perp,0}} = \Big(\frac{E_f}{E_0}\Big)_\text{r} \frac{\sqrt{1-\mu_f^2}}{\sqrt{1-\mu^2}} =1.
\end{equation}
It means that the acceleration of a reflected particle only causes increase of its momentum parallel to the magnetic field and 
thus decrease of its pitch angle.

\begin{figure}[ht]
\centering
   \includegraphics[width=8cm]{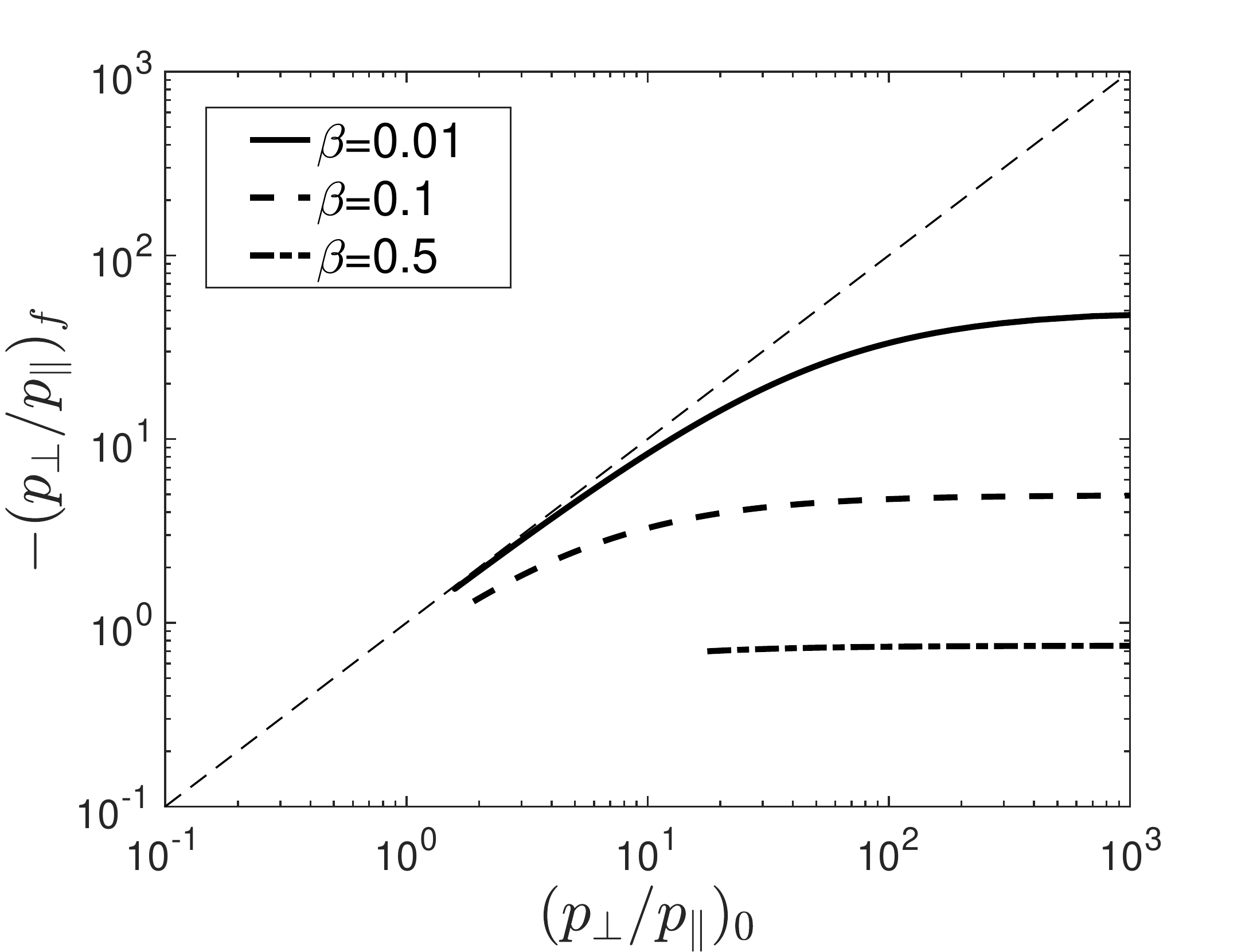}
\caption{The ratio of perpendicular to parallel momentum of a particle after reflection $-(p_\perp/p_\|)_f$
(with a negative sign added)
vs. that before reflection $(p_\perp/p_\|)_0$
at different $\beta$.
$\theta=45^\circ$ is used. 
The dashed line indicates the equalization between the two. }
\label{fig: cpp}
\end{figure}

The ratio of the perpendicular and parallel components of 
particle momentum is then 
\begin{equation}
   \Big(\frac{p_\perp}{p_\|}\Big)_f = \frac{\sqrt{1-\mu_f^2}}{\mu_f}. 
\end{equation}
In Fig. \ref{fig: cpp}, we compare the above ratio after reflection to its value before reflection 
$(p_\perp/p_\|)_0 = \sqrt{1-\mu^2}/\mu$ over the range of $\mu$ given in Eq. \eqref{eq: rmofrsym}, 
where we adopt $B_\text{out}/B_\text{in} \sim \cos \theta$ and $\theta = 45^\circ$. 
Naturally, we see that toward a larger $\beta$, the acceleration of reflected particles leads to a more anisotropic pitch angle distribution. 
This finding can explain the numerical result in 
\citet{Kow12}, 
where they found that particles are preferentially accelerated in the direction parallel to the magnetic field
during turbulent reconnection.

With the increase of anisotropy in pitch angle distribution, more particles would fall in the loss cone. 
The acceleration of unreflected particles acts to suppress the anisotropy and allows
the particles to leave the loss cone
(see Section \ref{ssec: unrefl}).
In addition, turbulent scattering also tends to isotropize the particle distribution.
These effects are important for suppressing the anisotropy in
pitch angle distribution and sustaining the mirror reflection. 
In our analysis through the paper, we will consider isotropic pitch angle distribution for simplicity.

\subsubsection{Pitch-angle averaged energy gain}

\begin{figure*}[ht]
\centering
\subfigure[]{
   \includegraphics[width=8cm]{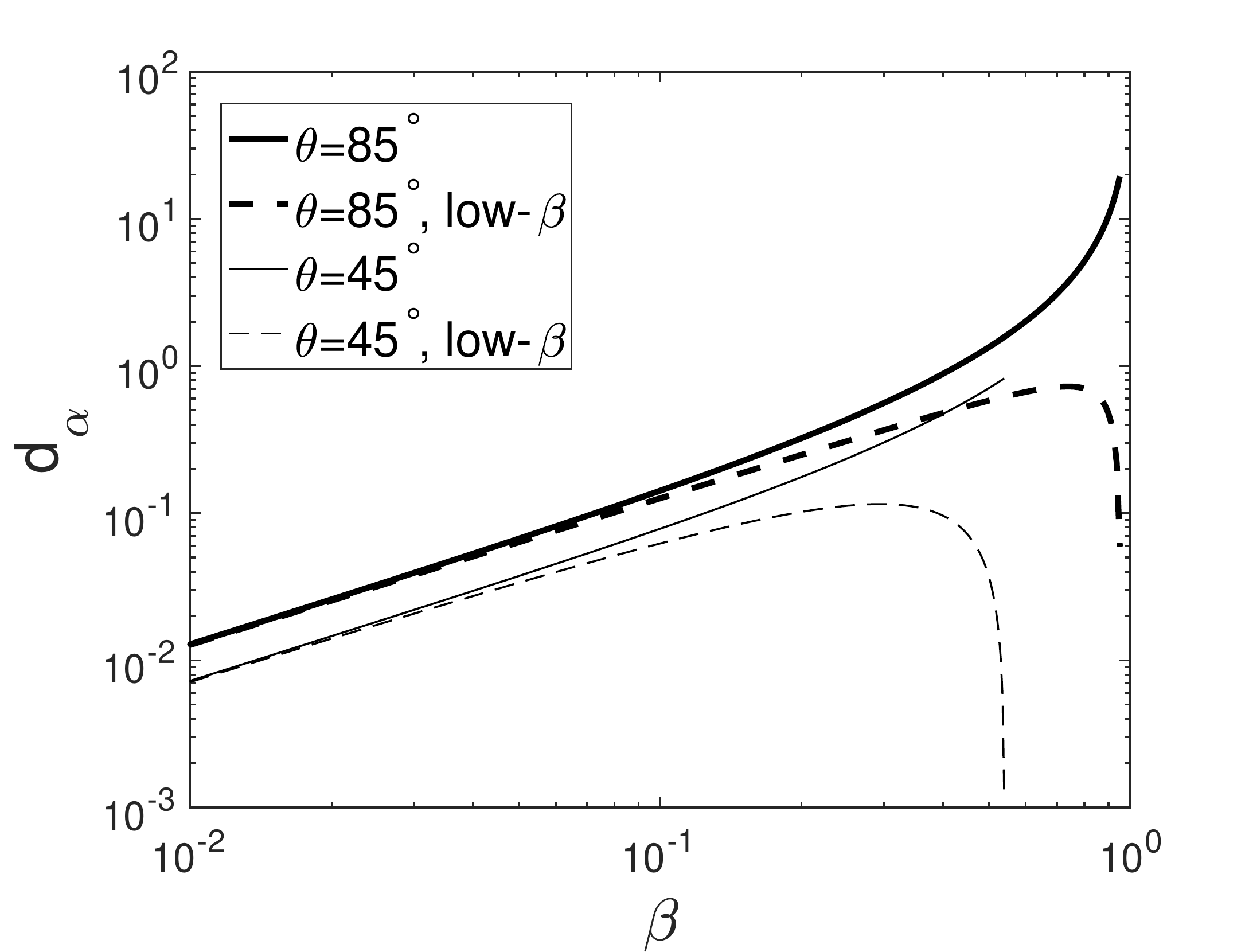}\label{fig: da}}
\subfigure[]{
   \includegraphics[width=8cm]{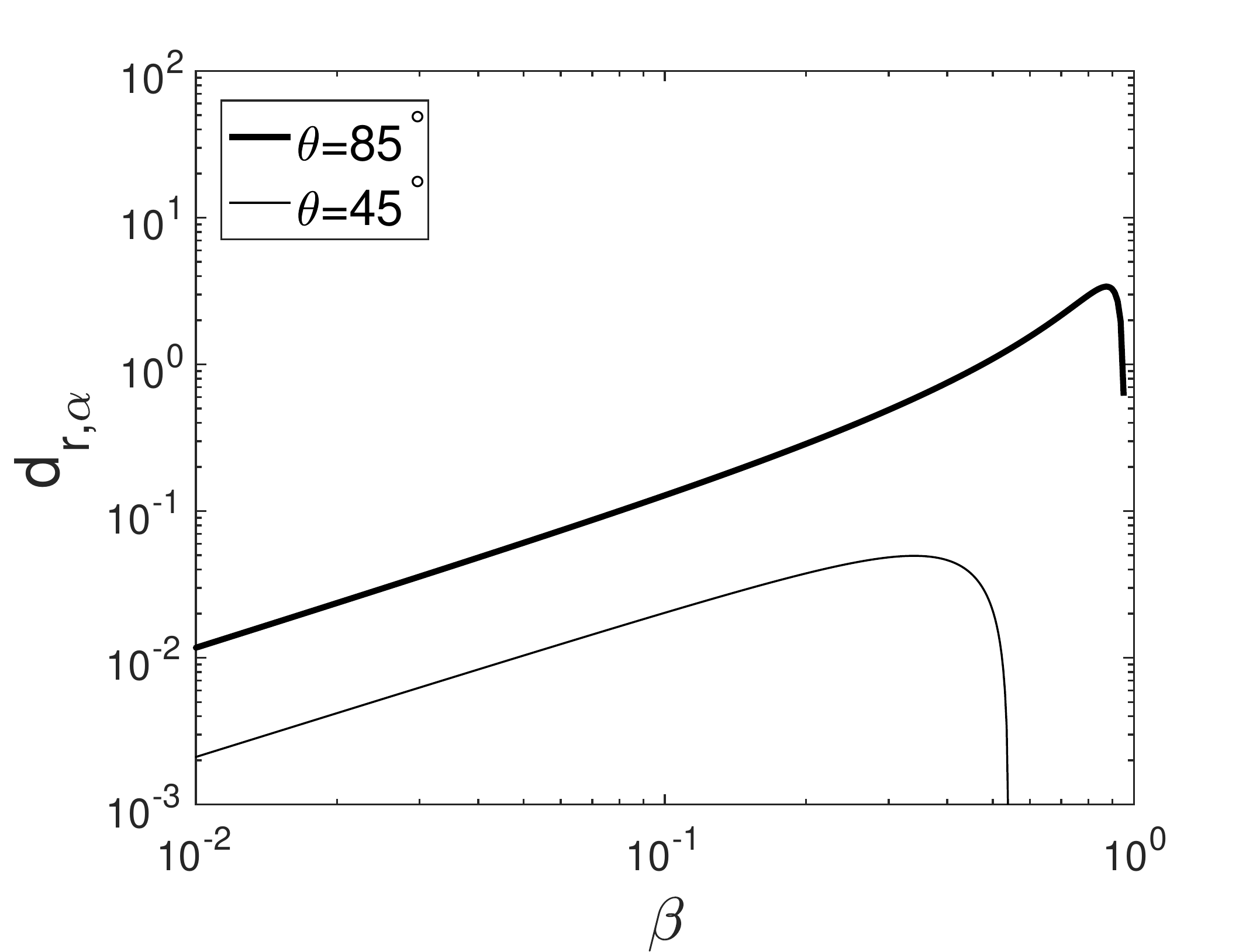}\label{fig: dra}}
\caption{
(a) Pitch-angle averaged fractional energy gain $d_\alpha$ as a function of $\beta$ for reflected particles.
The thick and thin solid lines are calculated using 
Eq. \eqref{eq: genda} for different $\theta$.
The dashed lines are their approximation in the low-$\beta$ limit given by Eq. \eqref{eq: sbeda}.
(b) Pitch-angle weighted fractional energy gain $d_{\text{r},\alpha}$ as a function of $\beta$ for reflected particles. 
The thick and thin solid lines are calculated using 
Eq. \eqref{eq:drarefpaw} for different $\theta$.
}
\label{fig: comdar}
\end{figure*}

The fractional energy gain $(E_f/E_0)_\text{r}-1$ of reflected particles has $\mu$ dependence (Eq. \eqref{eq: enf0}).
At fixed $f B_\text{in}/\sqrt{4\pi\rho}$,
for isotropic pitch angle distribution, we find 
the pitch-angle averaged fractional energy gain during reflection at given $\theta$ and inflow obliquity $\alpha$ as 
\begin{equation}\label{eq: genda}
   d_\alpha = \frac{\int V_\text{rel} \Big[\Big(\frac{E_f}{E_0}\Big)_\text{r} - 1\Big] d\mu }{\int V_\text{rel} d\mu },
\end{equation}
where $V_\text{rel}$ is the rate for particles to approach the inflow, 
\begin{equation}\label{eq: vrelpro}
   V_\text{rel} = \frac{\mu \cos\alpha + \frac{U_\text{in}}{c}}{1 + \frac{U_\text{in}\mu\cos\alpha}{c}},
\end{equation}
and the integral range is given by Eq. \eqref{eq: rmofrsym}.
A similar expression of $d_\alpha$ can be found in \citet{Ostr88} for studying the SDA at an oblique shock. 
By considering $V_\text{rel} \approx \mu\cos\alpha+ U_\text{in}/c$, we find approximately
\begin{equation}\label{eq:drappalathe}
   d_\alpha \approx \frac{2\beta [(\beta+\mu_c)^3 - \beta^3]}{3(\beta \mu_c + \frac{\mu_c^2}{2})(1-\beta^2)},
\end{equation}
where  
\begin{equation}
   \mu_c =  \frac{\sqrt{1-\frac{B_\text{out}}{B_\text{in}}}-\beta}{1 - \beta \sqrt{1- \frac{B_\text{out}}{B_\text{in}}}}.
\end{equation}
At a given $\theta$, $d_\alpha$ is only a function of $\beta$.
As shown in Fig. \ref{fig: da},
where we assume $B_\text{out}/B_\text{in} \sim \cos\theta$,  
$d_\alpha$ can significantly exceed unity when $\theta$ and $\beta$ are large. 
At a large $\theta$ with a weak guide field, 
due to the strong mirroring effect, 
the reflection can happen at a large $\beta$ (see Eq. \eqref{eq: conrfleb}), leading to a significant energy gain. 
In the low-$\beta$ limit, by using $(E_f /E_0)_\text{r} \approx 1+2\beta \mu$ and 
$V_\text{rel} \approx \mu \cos\alpha$, we find 
\begin{equation}\label{eq: sbeda}
    d_\alpha \approx \frac{4\beta}{3} \mu_c.
\end{equation}
It provides a good approximation when $\beta$ is small (see Fig. \ref{fig: da}). 

At given $\theta$ and $\alpha$, the probability of reflection is 
\citep{Ostr88}
\begin{equation}
   P_{\text{r},\alpha} = \frac{S_\text{r}}{S_\text{r} + S_\text{nr}}, 
\end{equation}
where 
\begin{equation}
    S_\text{r/nr} = \int V_\text{rel} d\mu
\end{equation}
with the integral range given by Eq. \eqref{eq: rmofrsym} for $S_\text{r}$ and Eq. \eqref{eq:loscmur}
for $S_\text{nr}$.
Therefore, we have the pitch-angle weighted fractional energy gain for reflection at given $\theta$ and $\alpha$ as 
\begin{equation}\label{eq:drarefpaw}
\begin{aligned}
    d_{\text{r},\alpha} &= P_{\text{r},\alpha} d_\alpha \\
                       &= \frac{\int_0^{\mu_c} V_\text{rel} \Big[\Big(\frac{E_f}{E_0}\Big)_\text{r}-1\Big]d\mu}{\int_0^1 V_\text{rel}d\mu} \\
                       &\approx \frac{2\beta [(\beta+\mu_c)^3-\beta^3]}{3(\beta+\frac{1}{2})(1-\beta^2)}.
\end{aligned}
\end{equation} 
As shown in Fig. \ref{fig: dra},
at the maximum $\beta$ for reflection, $\mu_c = 0$, and thus $d_{\text{r},\alpha}$ becomes $0$.
We find that $P_{\text{r},\alpha}$ is approximately equal to 
\begin{equation}
   P_{\text{r},\alpha} \approx \frac{\mu_c (\beta + \frac{\mu_c}{2})}{\beta+\frac{1}{2}},
\end{equation}
which can be reduced to $P_{\text{r},\alpha} \approx \mu_c^2$ at a small $\beta$.
At a given $\theta$, as the range of $\mu$ for reflection decreases with increasing $\beta$, 
$P_{\text{r},\alpha}$ becomes small at a large $\beta$, resulting in the drop of $d_{\text{r},\alpha}$ at the maximum $\beta$ for 
reflection.
At a given $\beta$, the range of $\mu$ for reflection increases with increasing $\theta$. 
When $\theta$ is large, 
$d_{\text{r},\alpha}$ is close to $d_\alpha$ except when $\beta$ approaches unity.

\subsection{Unreflected particles}
\label{ssec: unrefl}

The unreflected particles in the loss cone cannot penetrate deep into the inflow because of the strong incoming magnetic fields. 
After entering the inflow, they are attached to the incoming magnetic fields threading the inflow 
and advected back into the reconnection region. 
For a particle entering and exiting from the inflow, the ratio of the energies after ($E_f$) and before ($E_0$) the interaction is 
\begin{equation}\label{eq: unrfefe0}
   \Big( \frac{E_f}{E_0} \Big)_\text{ur}
    = \frac{1+2\beta_\text{in} \xi + \beta_\text{in}^2}{1-\beta_\text{in}^2},
\end{equation}
where $\beta_\text{in} = U_\text{in}/c$, and 
$\xi$ is the cosine of the angle $\Phi$ between the particle velocity and the inflow direction ($y$ direction).
The derivation of Eq. \eqref{eq: unrfefe0} is similar to that of Eq. \eqref{eq: enf0} for a reflected particle, but the Lorentz transformation is 
between the rest frames of the reconnection region and the inflow.
In the above calculation 
we assume that the particle enters the inflow with $\xi^\prime$  
and exits with $-\xi^\prime$, 
where $\xi^\prime$ in the rest frame of the inflow is related to $\xi$ by 
\begin{equation}
   \xi = \frac{\xi^\prime - \beta_\text{in}}{1-\beta_\text{in} \xi^\prime}.
\end{equation}
Although the form of Eq. \eqref{eq: unrfefe0} is similar to Eq. \eqref{eq: enf0} of a reflected particle,
the energy gain of an unreflected particle does not depend on the local obliquity of the inflow. 
As $\beta_\text{in} < \beta$, the energy gain of an unreflected particle after a single encounter with the inflow 
can be much smaller than that of a reflected particle. 
Moreover,
the return of an unreflected particle to the reconnection region is not due to the mirror reflection, but the advection of the inflow.

By performing the average over $\Phi$, we find the mean fractional energy gain of unreflected particles as 
\begin{equation}\label{eq: dunref}
    d_\text{ur} = \frac{\int_0^1 V_\text{rel,ur}  \Big[\Big(\frac{E_f}{E_0}\Big)_\text{ur}-1\Big]d\xi}{\int_0^1 V_\text{rel,ur} d\xi},
\end{equation}
where 
\begin{equation}
   V_\text{rel,ur} = \frac{\xi + \beta_\text{in}}{1+ \beta_\text{in} \xi}.
\end{equation}
Here we adopt the range $0<\xi<1$ for unreflected particles in the loss cone with different pitch angles with respect to the turbulent 
magnetic fields in the reconnection region. 
By taking $V_\text{rel,ur} \approx \xi+\beta_\text{in}$, we approximately have 
\begin{equation}\label{eq: durappe}
   d_\text{ur} \approx \frac{2\beta_\text{in} (3\beta_\text{in}^2 + 3 \beta_\text{in} + 1)}{3 (\beta_\text{in}+\frac{1}{2}) (1-\beta_\text{in}^2)} .
\end{equation}
When $\beta_\text{in}$ is small, it can be further simplified to
\begin{equation}\label{eq:dursimsb}
   d_\text{ur} \approx \frac{4 \beta_\text{in}}{3}.
\end{equation}

We note that the result in Eq. \eqref{eq:dursimsb}
is similar to the average fractional energy gain 
$(4/3) V/c$
after one round trip crossing and recrossing the shock in parallel shock acceleration, 
where $V=(3/4) U_\text{sh}$, and $U_\text{sh}$ is the shock speed 
\citep{Longairbook}.
Despite the similarity, in the case of shock acceleration, the energy gain comes from the kinetic energy of the converging flows
due to shock compression. 
A particle crossing the shock is scattered and isotropized within the moving downstream flow, and then returns to the upstream medium. 
For the reconnection acceleration of an unreflected particle, the energy gain comes from the kinetic energy of reconnection-driven inflows. 
The particle is coupled to the moving flow not by frequent scattering, but by attaching to the inflow magnetic field.

Following the similar analysis as that in Section \ref{ssec:ani}, 
we see that after an encounter of an unreflected particle with the inflow, $|\xi|$ increases (see Eq. \eqref{eq: mufinc}).
With respect to the turbulent magnetic field with a small $y$ component, 
this causes the increase of particle perpendicular momentum and pitch angle, leading to the suppression of the anisotropy 
caused by acceleration of reflected particles. 


\subsection{Averaged energy gain of reflected and unreflected particles}
\label{ssec: averefunr}

The reconnection-driven turbulence induces a range of local obliquities of the inflow with respect to the turbulent magnetic field 
${\bf B}_\text{out}$.
The range of pitch angles for reflection to happen depends on the local inflow obliquity (see Section \ref{sssec: enerefl}). 
Under the consideration of the acceleration of both reflected and unreflected particles, we define the averaged fractional energy gain for 
reconnection acceleration as 
\begin{equation}\label{eq:fintotd}
   d = P_\text{r} d_r
  + (1-P_\text{r} ) d_\text{ur},
\end{equation}
where 
\begin{equation}\label{eq: findr}
    d_\text{r} =  \frac{\int_{\cos(\alpha_\text{max})}^{\cos(\alpha_\text{min})} \int_0^{\mu_c} V_\text{rel} \Big[\Big(\frac{E_f}{E_0}\Big)_r-1\Big]d\mu d\cos\alpha}{\int_{\cos(\alpha_\text{max})}^{\cos(\alpha_\text{min})}\int_0^{\mu_c} V_\text{rel} d \mu d\cos\alpha} 
\end{equation}
is the pitch-angle and obliquity averaged fractional energy gain of reflected particles, 
\begin{equation}\label{eq: finpr}
   P_\text{r} =  \frac{\int_{\cos(\alpha_\text{max})}^{\cos(\alpha_\text{min})} \int_0^{\mu_c} V_\text{rel} d\mu d\cos\alpha}{\int_0^{\cos(\alpha_\text{min})}\int_0^1 V_\text{rel} d \mu d\cos\alpha}
\end{equation}
is the probability of reflection, 
and $d_\text{ur}$ as the averaged 
fractional energy gain of unreflected particles 
is given by Eq. \eqref{eq: dunref}.
At a given $\theta$, $d_r$ depends on the range of $\beta$. 
The smallest obliquity depends on the level of turbulence and 
is $\alpha_\text{min} = 90^\circ- \delta \alpha \sim 90^\circ - \theta$ (see Section \ref{ssec:turrec}).
As both the level of reconnection-driven turbulence and $U_\text{in}$ depend on $\theta$,
the corresponding $\beta$ is (Eq. \eqref{eq: velin})
\begin{equation}\label{eq: betamin}
        \beta_\text{min} \approx \frac{U_\text{in}}{c \cos (\pi/2 - \theta)}
                                  = \frac{f B_\text{in} }{c\sqrt{4\pi\rho}  },
\end{equation}
which is independent of $\theta$. 
The largest obliquity $\alpha_\text{max}$ allowed for reflection is limited by the condition in Eq. \eqref{eq: conrfleb}, where we assume $B_\text{out}/B_\text{in} \sim \cos\theta$. So the corresponding $\beta_\text{max}$ only depends on $\theta$. 
Toward a larger $\theta$, the mirroring effect becomes stronger, and thus 
the reflection can happen at larger $\beta$ values, leading to a larger $d_r$.
$d_\text{ur}$ only depends on $\beta_\text{in}$ (see Eq. \eqref{eq: durappe}) and thus also increases with $\theta$.
The first term in Eq. \eqref{eq:fintotd}
corresponding to the acceleration of reflected particles only exists when $\beta_\text{max}>\beta_\text{min}$, i.e., 
$\cos\theta < 1- \beta_\text{min}^2$.

When $\beta_\text{max}\gg \beta_\text{min}$, we approximately have 
\begin{equation}\label{eq: prnumtor}
\begin{aligned}
   & \int_{\cos(\alpha_\text{max})}^{\cos(\alpha_\text{min})} \int_0^{\mu_c} V_\text{rel} d\mu d\cos\alpha \\
    \approx & \frac{U_\text{in}^2}{c^2}\int_{\beta_\text{min}}^{\beta_\text{max}} \frac{1}{\beta^3}  \Big(\frac{\mu_c^2}{2}+\beta \mu_c\Big) d\beta \\
    \approx & \frac{\beta_\text{max}^2 U_\text{in}^2}{2 c^2} \int_{\beta_\text{min}}^{\beta_\text{max}} \frac{d\beta}{\beta^3}  \\
    \approx & \frac{\beta_\text{max}^2 U_\text{in}^2}{4c^2 \beta_\text{min}^2},
\end{aligned}
\end{equation}
and 
\begin{equation}
\begin{aligned}
  & \int_0^{\cos(\alpha_\text{min})}\int_0^1 V_\text{rel} d \mu d\cos\alpha \\
   \approx & \frac{U_\text{in}^2}{c^2} \int_{\beta_\text{min}}^\infty \frac{1}{\beta^3} \Big(\frac{1}{2}+\beta\Big) d\beta \\
    \approx & \frac{U_\text{in}^2}{4 c^2 \beta_\text{min}^2},
\end{aligned}
\end{equation}
where $\mu_c \approx \beta_\text{max} - \beta$ is used, and the integral is dominated by small $\beta$.
Therefore, we have 
\begin{equation}\label{eq:apppr}
   P_r \approx \beta_\text{max}^2 = 1 - \frac{B_\text{out}}{B_\text{in}},
\end{equation}
which only depends on $\theta$ and increases with $\theta$.
To derive the approximate expression of $d_r$, we also find 
\begin{equation}\label{eq: drnumtor}
\begin{aligned}
 &  \int_{\cos(\alpha_\text{max})}^{\cos(\alpha_\text{min})} \int_0^{\mu_c} V_\text{rel} \Big[\Big(\frac{E_f}{E_0}\Big)_r-1\Big]d\mu d\cos\alpha \\
  \approx & \frac{2 U_\text{in}^2}{3 c^2} \int_{\beta_\text{min}}^{\beta_\text{max}} \frac{(\mu_c+\beta)^3 -\beta^3}{\beta^2 (1-\beta^2)} d\beta \\
  \approx & \frac{2 U_\text{in}^2 \beta_\text{max}^3}{3 c^2 \beta_\text{min}},
\end{aligned}
\end{equation}
where we consider that the integral is dominated by small $\beta$. 
Combining Eqs. \eqref{eq: prnumtor} and \eqref{eq: drnumtor} yields 
\begin{equation}\label{eq:appdr}
   d_r \approx \frac{8\beta_\text{min} \beta_\text{max}}{3}.
\end{equation}
Together with the approximate expression of $d_\text{ur}$ at a small $\beta_\text{in}$ in Eq. \eqref{eq:dursimsb}, 
we obtain 
\begin{equation}\label{eq:apptotd}
   d \approx \frac{8\beta_\text{min} \beta_\text{max}^3}{3} + \frac{4 (1-\beta_\text{max}^2) \beta_\text{in}}{3} .
\end{equation}
It can be further simplified to 
\begin{equation}\label{eq:sthedsim}
   d\approx d_\text{ur} \approx \frac{4\beta_\text{in}}{3}
\end{equation}
at a small $\theta$, and 
\begin{equation}\label{eq:lthedsim}
   d\approx d_\text{r} \approx \frac{8\beta_\text{min}}{3}
\end{equation}
at a large $\theta$.

\begin{figure*}[ht]
\centering
\subfigure[]{
   \includegraphics[width=8cm]{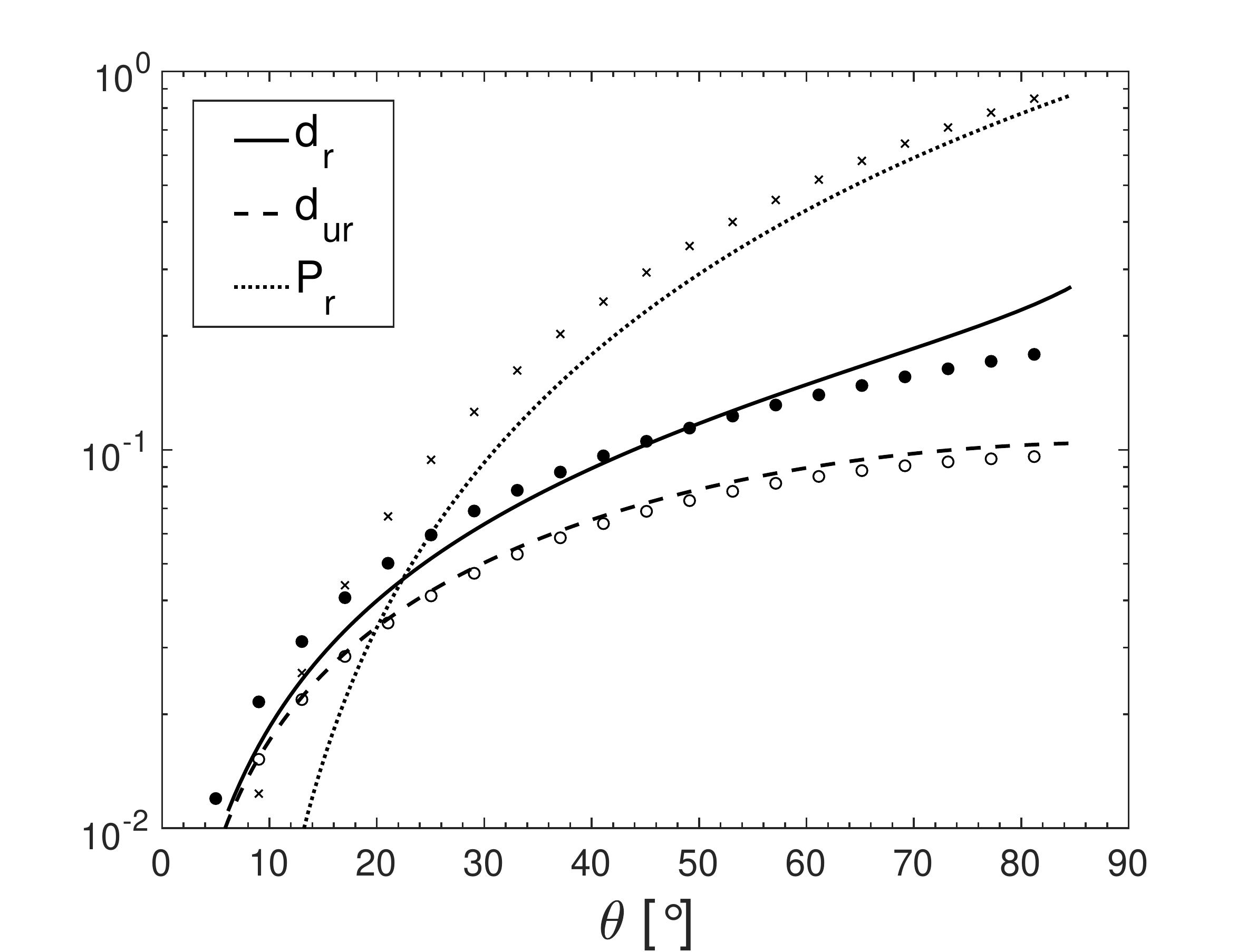}\label{fig: drdur}}
\subfigure[]{
   \includegraphics[width=8cm]{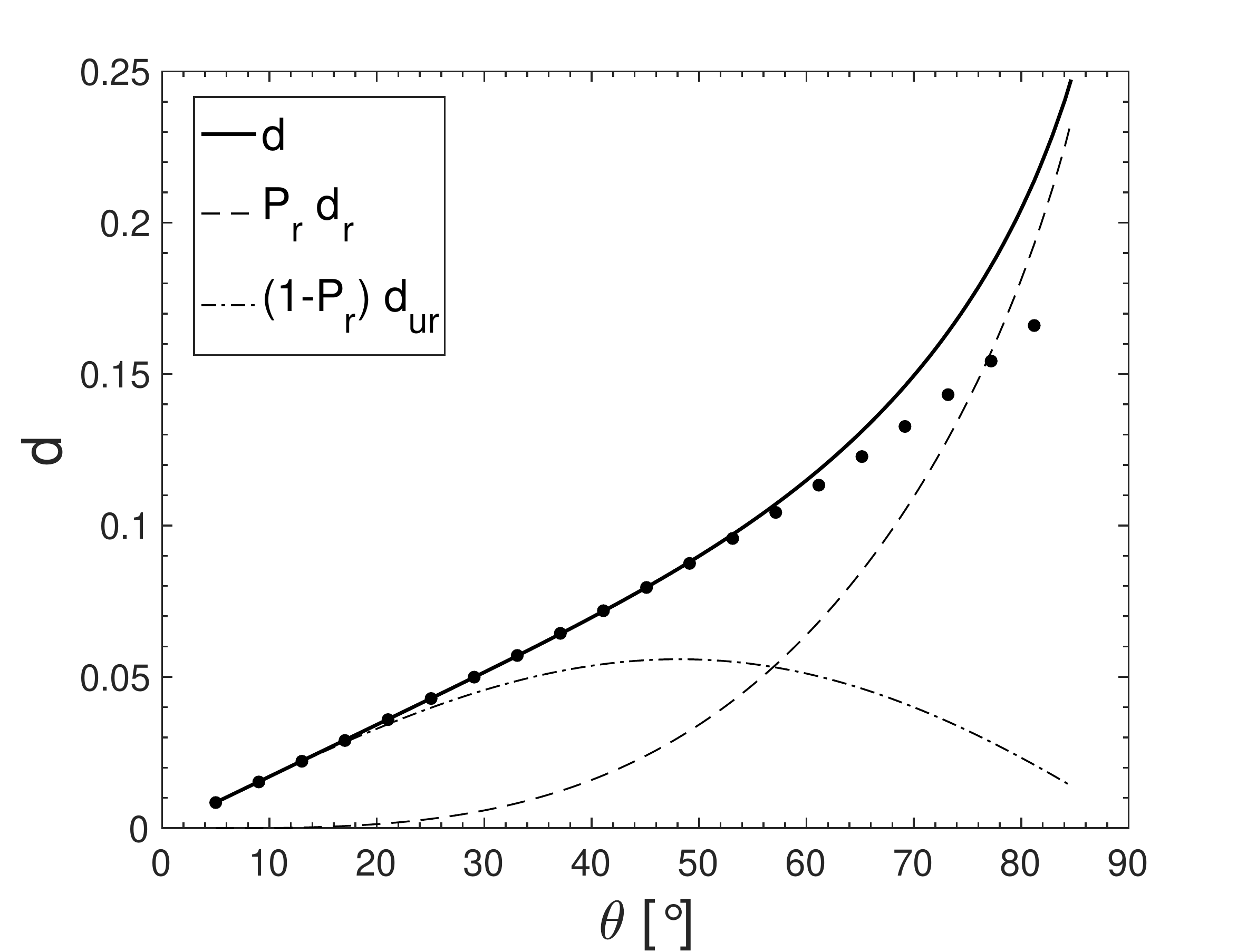}\label{fig: daves}}
\subfigure[]{
   \includegraphics[width=8cm]{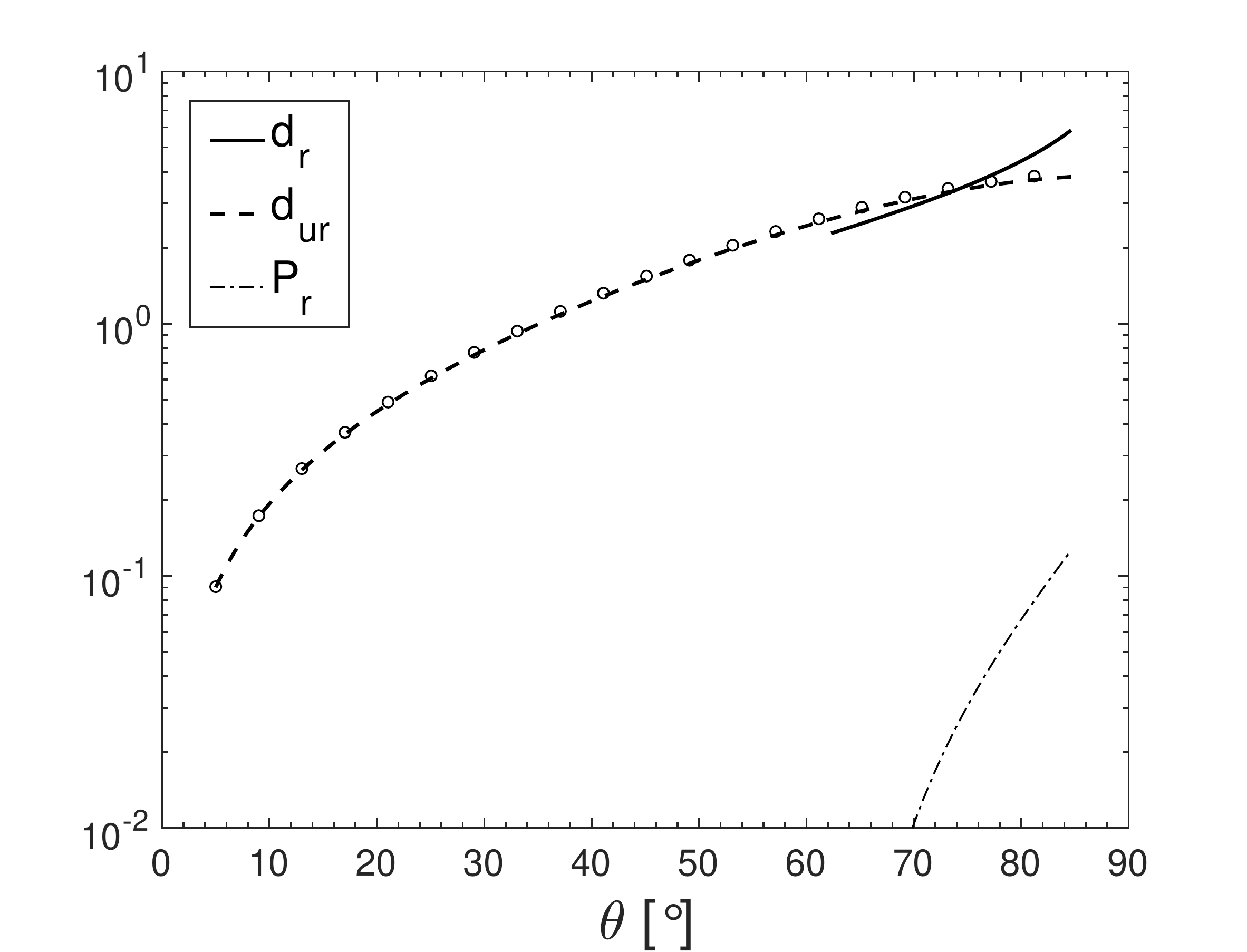}\label{fig: drdur2}}
\subfigure[]{
   \includegraphics[width=8cm]{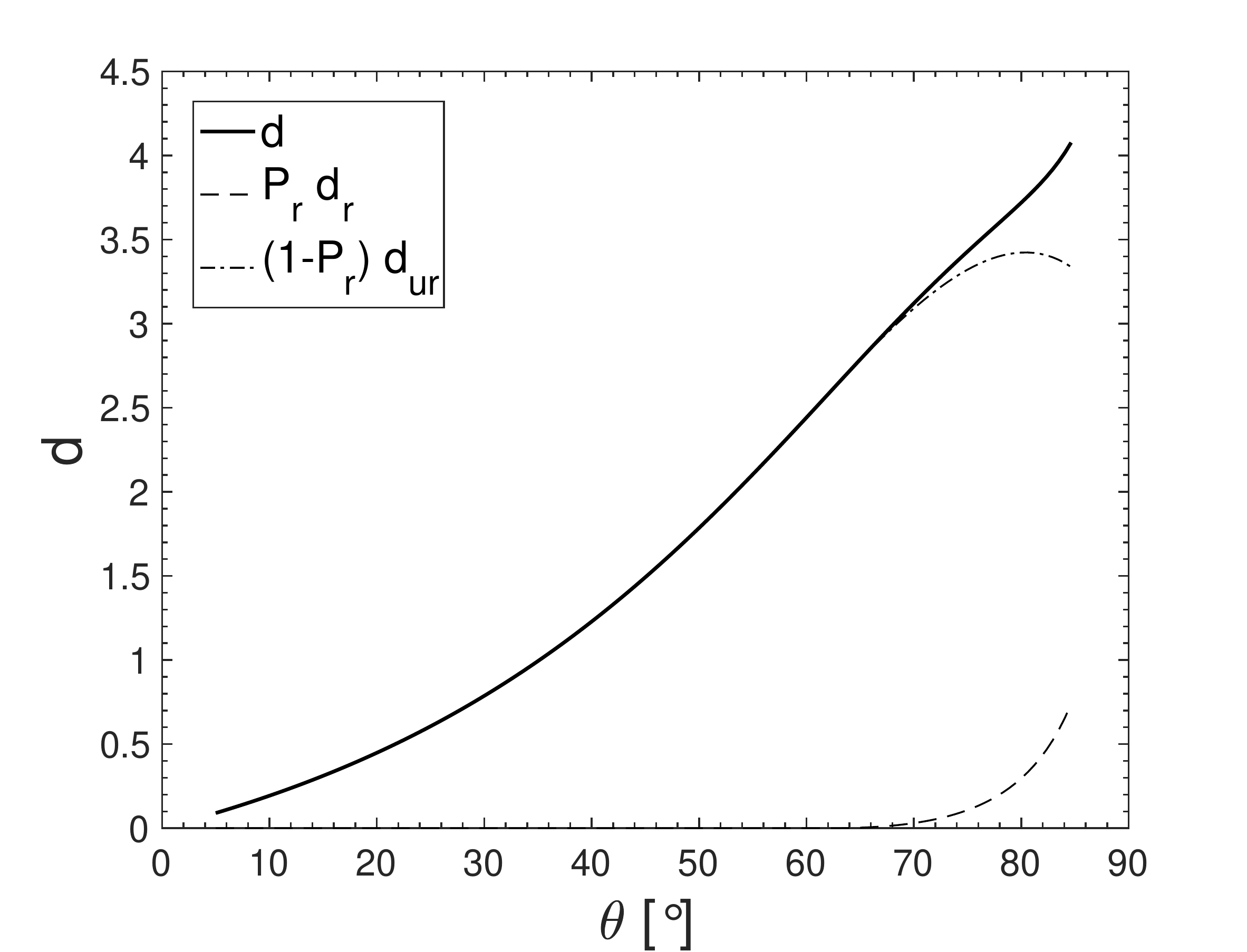}\label{fig: davel}}
\caption{(a) Averaged fractional energy gain of reflected particles $d_\text{r}$ (Eq. \eqref{eq: findr}) and unreflected particles $d_\text{ur}$ (Eq. \eqref{eq: dunref}).
$P_r$ (Eq. \eqref{eq: finpr}) is the probability of reflection. 
Filled circles, open circles, and crosses correspond to the approximate expressions of $d_\text{r}$ (Eq. \eqref{eq:appdr}), $d_\text{ur}$ (Eq. \eqref{eq:dursimsb}), and $P_\text{r}$ (Eq. \eqref{eq:apppr}) in the limit of $\beta_\text{min}\ll\beta_\text{max}$. 
(b) Total averaged fractional energy gain $d$ (Eq. \eqref{eq:fintotd}) during reconnection acceleration.
The filled circles correspond to the approximate expression of $d$ (Eq. \eqref{eq:apptotd}).
(a) and (b) correspond to $fB_\text{in}/(c\sqrt{4\pi\rho}) = 0.07$.
(c) and (d) are the same as (a) and (b), but for $fB_\text{in}/(c\sqrt{4\pi\rho}) = 0.7$.
The open circles in (c) correspond to $d_\text{ur}$ given by Eq. \eqref{eq: durappe}.
}
\label{fig: dave}
\end{figure*}

In Figs. \ref{fig: drdur} and \ref{fig: daves}, 
we present $d_\text{r}$, $d_\text{ur}$, and $d$ calculated at $fB_\text{in}/(c\sqrt{4\pi\rho}) = 0.07$.
At a small $\theta$, $d_\text{r}$ and $d_\text{ur}$ are comparable, 
but $P_r$ is small because of the small ranges of obliquities and 
pitch angles allowed for reflection. 
Therefore, $d$ is dominated by $d_\text{ur}$.
With the increase of $\theta$, $P_r$ increases and reaches a large value at a large $\theta$.
Hence $d$ is dominated by $d_\text{r}$, which is larger than $d_\text{ur}$ due to the larger energy gain at a larger $\beta$ for 
reflection acceleration. 
We also present the approximate results in the limit of $\beta_\text{min} \ll \beta_\text{max}$ derived above. 
At a small $\theta$ with a small $\beta_\text{max}$, the condition $\beta_\text{min} \ll \beta_\text{max}$ is not well satisfied. 
At a large $\theta$, the consideration that the integral in Eq. \eqref{eq: drnumtor} is dominated by small $\beta$ is not well justified. 
Therefore, we see some deviations in $d_r$, $P_r$, and $d$ despite the generally good agreement.

In Figs. \ref{fig: drdur2} and \ref{fig: davel}, we present $d_\text{r}$, $d_\text{ur}$, and $d$ calculated at $fB_\text{in}/(c\sqrt{4\pi\rho}) = 0.7$. 
Given a larger $\beta_\text{min}$, the condition for reflection cannot be satisfied over a large range of $\theta$ with $\beta_\text{min}>\beta_\text{max}$,
where the energy gain is solely contributed by acceleration of unreflected particles with $d=d_\text{ur}$.
At a large $\theta$,
large $\beta$ values with a large $\beta_\text{min}$ give rise to a large $d_\text{r}$.
$d_\text{ur}$ also becomes significantly large with a large $\beta_\text{in}$.
As the range of pitch angles for reflection is limited at a large $\beta$, $P_r$ is small even at a large $\theta$, 
and thus $d$ is dominated by $d_\text{ur}$.
$d_\text{ur}$ is 
well approximated by Eq. \eqref{eq: durappe} (see Fig. \ref{fig: drdur2}).

We see that $d_\text{r}$ is in general larger than $d_\text{ur}$ when reflection can happen. 
However, sufficiently strong mirroring and moderate $\beta$ are required for a large $P_r$ and for $d$ to be dominated by $d_\text{r}$.
As a result, $d$ is usually dominated by $d_\text{ur}$ at a small $\theta$.
When $\beta_\text{min}$ is large, 
$d$ is basically determined by $d_\text{ur}$ over all $\theta$. 
By comparing our result with that in \citet{DeG05}, 
where they derived $(4/3) U_\text{in}/c$ as the pitch-angle averaged fractional energy gain per crossing of the reconnection region, 
we find that their expression only applies to unreflected particles in the case with $U_\text{in} \ll c$.

\section{Probability of escape and particle energy spectrum}
\subsection{Probability of escape}

Under the assumption of isotropic particle distribution in the local fluid frames, 
in the reference frame of the reconnection region (hereafter R frame), we find 
the number of particles that enter the reconnection region per unit time as
\begin{equation}\label{eq: genton}
\begin{aligned}
   &~~~~~~L_x L_z \int c \cos\phi f d^3 {\bm p}  \\
      &= 2\pi L_x L_z \int c \cos\phi f p^2 dp d cos\phi \\
      &= 2\pi c L_x L_z \int_0^\infty g {p^\prime}^2dp^\prime \int_0^1 \frac{\cos \phi d\cos\phi}{\Gamma_\text{in}^3 (1-\beta_\text{in}\cos\phi)^3} \\
      &= \pi c L_x L_z \int_0^\infty g {p^\prime}^2dp^\prime \frac{1}{ \Gamma_\text{in}^3 (1-\beta_\text{in})^2}\\
      &= \frac{n_\text{in}c}{4} L_x L_z \frac{1}{ \Gamma_\text{in}^3 (1-\beta_\text{in})^2} \\
      &=  \frac{n_Rc}{4} L_x L_z \frac{1}{ \Gamma_\text{in}^4 (1-\beta_\text{in})^2} ,
\end{aligned}
\end{equation}
where $\beta_\text{in} = U_\text{in}/c$, $\Gamma_\text{in} = 1/\sqrt{1-\beta_\text{in}^2}$, 
$n_\text{in}$ is the number density of particles in the inflow frame, 
$n_R = \Gamma_\text{in} n_\text{in}$ is the number density of particles in the R frame, 
$g$ is the distribution function in the inflow frame, which is equal to 
the distribution function $f$ in the R frame, 
$p^\prime$ and $p$ are particle momentum in the fluid frame and R frame, 
$\phi$ is the angle between ${\bf p}$ and the inflow direction in the R frame, 
and $L_z$ is the length along the $z$ direction. 
Here we apply the Lorentz transformation between the inflow frame and the R frame with 
$p^\prime = \Gamma_\text{in} p(1-\beta_\text{in} \cos\phi)$
following the approach in
\citet{NaTa95}, 
and the particle distribution 
becomes anisotropic in the R frame. 
When $\beta_\text{in}\ll1$, the above expression is approximately equal to 
\begin{equation}
     \frac{n_\text{in}c}{4} L_x L_z.
\end{equation}

Particles can escape from the reconnection region by gyrating around the turbulent guide field, which is advected 
away by the outflow from the reconnection region.  
Of the particles that enter the reconnection region, the number of escaping particles per unit time in the R frame is 
\begin{equation}\label{eq: genesn}
\begin{aligned}
      &~~~~~~\Delta L_z \int c \cos\phi f d^3 {\bm p} \\
      &= 2\pi c \Delta L_z \int_0^\infty g {p^\prime}^2dp^\prime \int_{-1}^1 \frac{\cos \phi d\cos\phi}{\Gamma_\text{out}^3 (1-\beta_\text{out}\cos\phi)^3} \\
      &= 4\pi c \Delta L_z \int_0^\infty g {p^\prime}^2dp^\prime  \Gamma_\text{out} \beta_\text{out} \\
      & = n_\text{out} c \Delta L_z \Gamma_\text{out} \beta_\text{out} \\
      & = n_R c \Delta L_z  \beta_\text{out},
\end{aligned}
\end{equation}
where $\beta_\text{out} = U_\text{out}/c$, $\Gamma_\text{out} = 1/\sqrt{1-\beta_\text{out}^2}$,
and $n_\text{out}$ is the number density of particles in the outflow frame. 
Similar to the above calculation, here $g$ and $p^\prime$ are the distribution function and particle momentum in the outflow frame, and $\phi$ is the angle between ${\bf p}$ and the outflow direction in the R frame. 
When $\beta_\text{out}\ll1$, the above expression approximately becomes 
\begin{equation}
   n_\text{out} U_\text{out} \Delta L_z. 
\end{equation}

The escape probability is the ratio of Eq. \eqref{eq: genesn} to Eq. \eqref{eq: genton},
\begin{equation}
   P_\text{esp} 
                       = \frac{\Delta}{L_x} \frac{4\beta_\text{out} }{(1+\beta_\text{in})^2}.
\end{equation}
As $\Delta /L_x = f$ and $\beta_\text{in} = f \beta_\text{out}$, we have 
\begin{equation}\label{eq: pesp}
   P_\text{esp} = \frac{4\beta_\text{in}}{(1+\beta_\text{in})^2},
\end{equation}
which is approximately 
\begin{equation}\label{eq:apppes}
  P_\text{esp} \approx 4 \beta_\text{in}
\end{equation}
when $\beta_\text{in} \ll1$. 
The probability for a particle to remain in the reconnection region is 
\begin{equation}\label{eq:probres}
  P = 1 - P_\text{esp} = 1- \frac{4\beta_\text{in}}{(1+\beta_\text{in})^2},
\end{equation}
which is small when $\beta_\text{in}$ is large. 
Fig. \ref{fig: pesp} shows $P_\text{esp}$ as a function of $\theta$. 
We see that with the increase of $fB_\text{in}/\sqrt{4\pi\rho}$ and $\theta$, $P_\text{esp}$ can approach unity at a sufficiently large $\beta_\text{in}$.


\subsection{Spectral index of accelerated particles}

\begin{figure*}[ht]
\centering
\subfigure[]{
   \includegraphics[width=8cm]{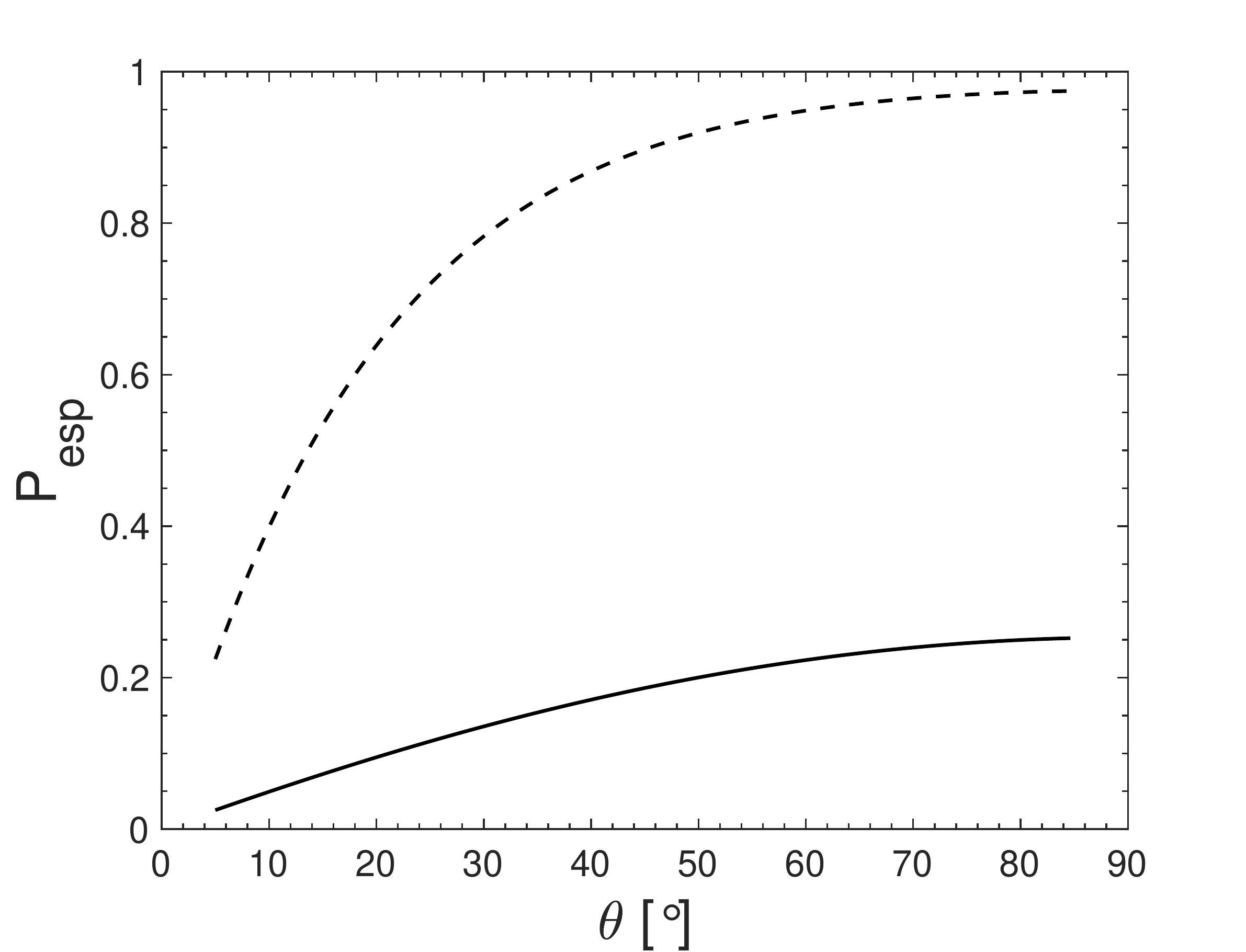}\label{fig: pesp}}
\subfigure[]{
   \includegraphics[width=8cm]{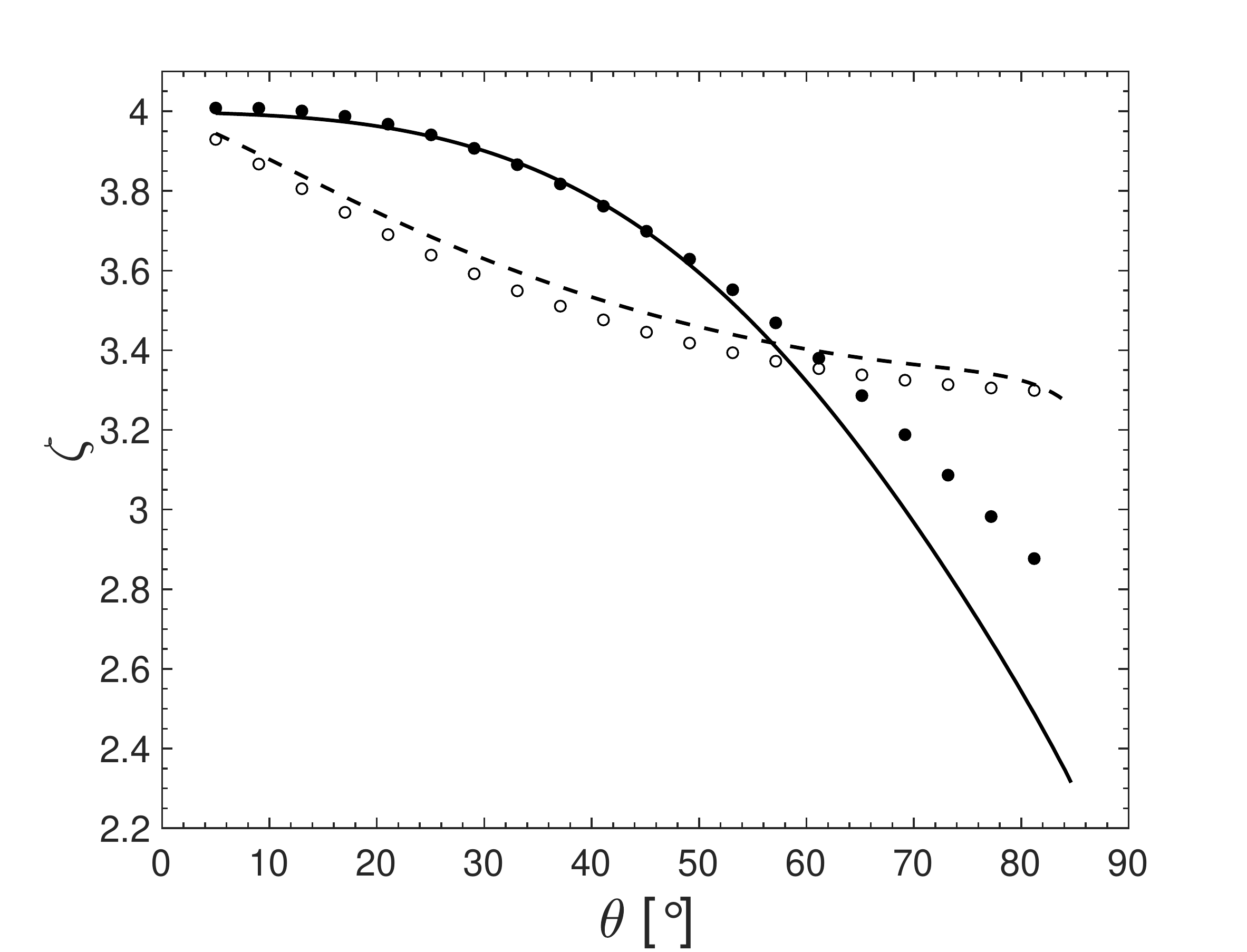}\label{fig: zet}}
\caption{ (a) $P_\text{esp}$ calculated using Eq. \eqref{eq: pesp}.
Solid and dashed lines correspond to $fB_\text{in}/(c\sqrt{4\pi\rho}) = 0.07$ and $fB_\text{in}/(c\sqrt{4\pi\rho}) = 0.7$, respectively.
(b) Spectral index $\zeta$ (Eq. \eqref{eq: spindx}) as a function of $\theta$. 
The same symbols as for (a) are used. 
Filled circles correspond to the approximate result with $d$ estimated by Eq. \eqref{eq:apptotd}.
Open circles correspond to the approximate result using $d \approx d_\text{ur}$, where $d_\text{ur}$ is given by Eq. \eqref{eq: durappe}.
}
\label{fig: pespzet}
\end{figure*}

The shape of particle energy spectrum 
$N(E) \propto E^{-\zeta}$, where $E$ is the particle energy, 
is determined by the competition between acceleration and escape of particles. 
It has the power-law index as 
\citep{Longairbook}
\begin{equation}\label{eq: spindx}
    \zeta = 1-\frac{\ln P}{\ln \epsilon},
\end{equation}
where $\epsilon = 1 + d$, $d$ is given by Eq. \eqref{eq:fintotd}, 
and $P$ is given by Eq. \eqref{eq:probres}.
Fig. \ref{fig: zet} shows $\zeta$ as a function of $\theta$ at different values of $fB_\text{in}/\sqrt{4\pi\rho}$.
We see that the reconnection-accelerated particles have a non-universal energy spectral index. 
At a small $\theta$ and a small $U_\text{in}$, the energy gain mainly comes from the acceleration of unreflected particles, and $P_\text{esp}$
is small. 
There is approximately 
(Eqs. \eqref{eq:sthedsim} and \eqref{eq:apppes}), 
\begin{equation}
  \zeta \approx 1 - \frac{\ln (1- 4\beta_\text{in})}{\ln(1+ \frac{4\beta_\text{in}}{3})} \approx 4,
\end{equation}
as shown in Fig. \ref{fig: zet}. This corresponds to the steepest energy spectrum of reconnection-accelerated particles. 
With the increase of $\theta$, $\zeta$ decreases because of the larger energy gain toward larger $\theta$. 
$\zeta$ at a large $\theta$ depends on the value of $U_\text{in}$.
When $U_\text{in}$ is large, although the energy gain is significant, 
the large $P_\text{esp}$ leads to a steep spectrum. 
When $U_\text{in}$ is small (relative to $c$), the energy gain is mainly contributed by the acceleration of reflected particles, and $P_\text{esp}$ is small.
We approximately have 
(Eqs. \eqref{eq:lthedsim} and \eqref{eq:apppes})
\begin{equation}
    \zeta \approx 1 - \frac{\ln (1- 4\beta_\text{in})}{\ln(1+ \frac{8\beta_\text{min}}{3})} \approx 2.5.
\end{equation}
This estimate is a bit  
larger than $\zeta$ at a large $\theta$ calculated for $fB_\text{in}/(c\sqrt{4\pi\rho}) = 0.07$ in Fig. \ref{fig: zet} due to 
the underestimate of $d$ (see Fig. \ref{fig: daves}).

Based on the above analysis, we see that $\zeta$ of reconnection-accelerated particles 
depends on $U_\text{in}$.
It approximately falls in the range 
$2.5 \lesssim \zeta < 4$.


\section{Acceleration time} 

As scattering diffusion of particles is not a necessary process for confining particles,
particles can ballistically move along magnetic field lines in the reconnection region.
As particles cannot penetrate deep into the inflows, we consider the crossing time of a particle as its 
residence time in the reconnection region. 
For isotropic pitch angle distribution, the crossing time of a particle is estimated as 
\begin{equation}\label{eq: gentau}
  \tau = \frac{\Delta}{c \langle \mu \rangle \langle \cos\alpha \rangle},
\end{equation}
where 
\begin{equation} 
   \langle \cos\alpha \rangle = \frac{\int_0^{\cos(\alpha_\text{min})} \cos\alpha d\cos\alpha}{\int_0^{\cos(\alpha_\text{min})} d\cos\alpha}
   \approx \frac{\sin\theta}{2},
\end{equation}
and 
\begin{equation}
    \langle \mu \rangle = \frac{\int_0^1 \mu d\mu}{\int_0^1 d\mu} = \frac{1}{2}.
\end{equation}
The turbulence-perturbed magnetic fields in the reconnection region have their orientations deviate from the original guide field direction. 
With the increase of $\theta$, a higher level of turbulence introduces a larger range of magnetic field orientations. 
As a result, $\tau$ decreases with increasing $\theta$.

\begin{figure}[ht]
\centering
   \includegraphics[width=8cm]{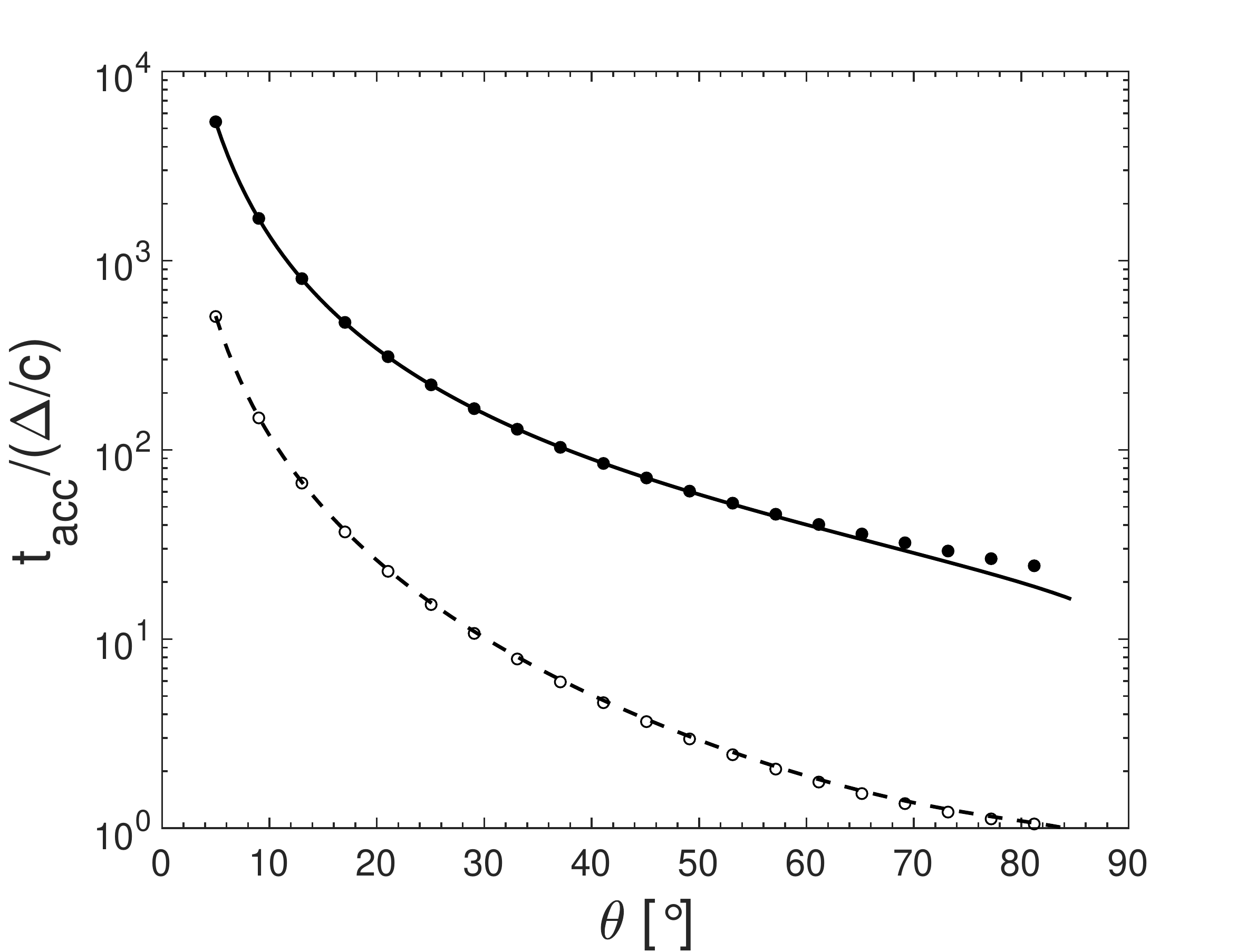}  
\caption{$t_\text{acc}$ (normalized by $\Delta/c$) vs. $\theta$ calculated using 
Eq. \eqref{eq:taccf}. 
Solid and dashed lines correspond to $fB_\text{in}/(c\sqrt{4\pi\rho}) = 0.07$ and $fB_\text{in}/(c\sqrt{4\pi\rho}) = 0.7$, respectively.
Filled circles correspond to the approximate result with $d$ estimated by Eq. \eqref{eq:apptotd}.
Open circles correspond to the approximate result using $d \approx d_\text{ur}$, where $d_\text{ur}$ is given by Eq. \eqref{eq: durappe}.
}
\label{fig: tacc}
\end{figure}

The acceleration time is then
\begin{equation}\label{eq:taccf}
    t_\text{acc} = \frac{\tau}{d},
\end{equation}
where $d$ is given by Eq. \eqref{eq:fintotd}.
At a small $\theta$, $d$ is dominated by $d_\text{ur}$. By using the approximate expression of $d_\text{ur}$
(Eq. \eqref{eq:dursimsb}) at a small $\beta_\text{in}$, we find 
\begin{equation}
   t_\text{acc} \approx \frac{3\Delta}{ \sin\theta U_\text{in}}, ~(\text{small}~ \theta).
\end{equation}
With the increase of $\theta$, $\tau$ decreases and $d$ increases. So the resulting $t_\text{acc}$ decreases. 
At a large $\theta$, when $U_\text{in}$ is small (relative to $c$), 
$d$ is dominated by $d_\text{r}$. By using the approximate expression of $d_\text{r}$ (Eq. \eqref{eq:lthedsim}),
we find
\begin{equation}
   t_\text{acc} \approx \frac{3\Delta}{2c \beta_\text{min}} \approx \frac{3\Delta}{2 U_\text{in}}, ~(\text{large}~ \theta, ~\text{small} ~\beta_\text{in}).
\end{equation}
As shown in Fig. \ref{fig: tacc}, at a large $\theta$ and a large $U_\text{in}$, 
$t_\text{acc}$ is close to the light crossing time of $\Delta$,
which has an approximate expression 
\begin{equation}
    t_\text{acc} \approx \frac{4\Delta}{c d_\text{ur}},
    ~(\text{large}~ \theta, ~\text{large} ~\beta_\text{in}),
\end{equation}
where $d_\text{ur}$ is given by Eq. \eqref{eq: durappe}.

{ In the above estimate of $\tau$, 
we consider that particles can ballistically cross the reconnection region with inefficient scattering by turbulence, 
as suggested by the measured particle trajectories in turbulent reconnection simulations 
\citep{Zjf22}.
We note that 
the turbulent superdiffusion perpendicular to the mean guide field 
(e.g., \citealt{XY13,Hucr22})
can shorten $\tau$ at a small $\theta$, while the nonresonant interaction with turbulent magnetic compressions can cause slow diffusion 
\citep{LX21}
and a larger $\tau$ at a large $\theta$. }

\section{Discussion}

\subsection{Comparison between reconnection acceleration and oblique shock acceleration}

At an oblique shock, particles are energized by a combination of SDA and DSA
\citep{Jok82,Deck88,Ostr88,Kirk89,XL22}.
{ They both have the same nature of Fermi acceleration via bouncing between
approaching magnetic mirrors, 
including the mirror created by 
shock compression or the ``mirrors" created by amplified magnetic irregularities upstream and downstream of the shock. }
The increase of shock obliquity causes a stronger shock compression of magnetic fields and thus a stronger mirroring effect, 
as well as a larger fractional energy gain.
Isotropic particle distribution and particle diffusion are needed for the particles to repeatedly return to the shock in both upstream and 
downstream regions. 
The isotropic particle distribution is traditionally related to efficient pitch-angle scattering without specifying the scattering process. 
It can also be realized when particles follow isotropically distributed turbulent magnetic fields in super-Alfv\'{e}nic turbulence 
\citep{XL22}, 
which are amplified by nonlinear turbulent dynamo 
\citep{XL16,XuL17,Hu22sh}.

For reconnection acceleration, 
particles undergo head-on collisions with the inflows driven by magnetic reconnection. 
Unlike the shock acceleration, 
the compression of gas and 
{ density change}
does not play an important role for the reconnection acceleration.
The mirror reflection of particles is caused by the 
annihilation of the antiparallel components of magnetic fields. 
The resulting change of magnetic field strength is not limited by the gas compression ratio as in the case of shock acceleration.
At a large $\theta$ with a weak guide field, a much stronger mirroring effect than that of a highly oblique shock is expected.
Therefore, when the condition for reflection is satisfied, a single encounter with the inflow can 
result in a much larger fractional energy gain compared to the interaction with an oblique shock. 
Unlike the shock acceleration, particles with loss-cone pitch angles are not transmitted through the inflow. 
The unreflected particles are attached to the incoming magnetic fields and sent back to the 
reconnection region 
as the incoming magnetic fields reconnect. 
They gain energy during this process. 

Compared with nonrelativistic shock acceleration,
it becomes much more challenging for particles to return to the shock from downstream 
for relativistic shock acceleration.
For relativistic reconnection acceleration, 
the return of particles to the reconnection region occurs naturally, allowing efficient acceleration.

Both magnetic mirroring and convergence of magnetic fields contribute to the confinement of particles in the reconnection region. 
Therefore, scattering diffusion is not a necessary confining mechanism for reconnection acceleration. 
For the shock acceleration, the anisotropic diffusion of particles is important for determining the acceleration efficiencies at different 
shock obliquities 
\citep{Jok87,XL22}. 
For the reconnection acceleration, instead of having a slow diffusive motion, particles can ballistically move along magnetic field lines
in the reconnection region
{ when the scattering of particles by turbulence is inefficient}, 
leading to a much shorter time between successive accelerations than that for diffusive shock acceleration. 
Both the significant energy gain and short residence time at a large $\theta$ can result in much shorter acceleration timescale 
and more efficient acceleration than shock acceleration. 

The escape of particles in the far downstream region of a shock is caused by 
the advection of particles with the flow due to isotropic particle distribution. 
The escape of particles from the reconnection region is caused by the advection of the guide field with the outflow, irrespective of 
the particle distribution. 
For a highly oblique shock, $P_\text{esp}$ becomes large due to the large fluid speed along the magnetic field. 
For reconnection acceleration, at a large $\theta$, $P_\text{esp}$ becomes large when the inflow speed (relative to $c$) is large.
Both shock acceleration and reconnection acceleration have non-universal indices of 
particle energy spectra. 
Acceleration at a quasi-parallel shock with a small shock speed leads to a spectral index equal to $2$
\citep{Bell78}.
A steep particle energy spectrum with the spectral index much larger than $2$ is expected at a large shock obliquity and a large shock speed
\citep{XL22,Xu22}.
For reconnection acceleration, 
we find the particle spectral index dependent on $U_\text{in}$ and $\theta$ and 
falling in the range $2.5 \lesssim \zeta < 4$.

For both shock acceleration and reconnection acceleration, turbulence plays a critical role. 
For the former case, the amplification of magnetic fields via the nonlinear turbulent dynamo is important for confining particles near the shock 
and determining the largest achievable particle energy. 
Turbulence introduces variations of shock obliquities, enables the combined acceleration mechanism of SDA and DSA, 
and affects the diffusion of particles in directions parallel and perpendicular to magnetic fields and acceleration efficiency
\citep{XL22}. 
For the latter case, turbulence enables fast magnetic reconnection. The efficient magnetic reconnection is the prerequisite for efficient 
reconnection acceleration. 
Turbulence introduces variations of inflow obliquities, enables the RDA of reflected particles, 
and affects the inflow speed and acceleration efficiency.


{ \citet{Druf12}
applied the ``box" model for shock acceleration
\citep{Drur99}
to reconnection acceleration.
For the reconnection model, 
\citet{Druf12}
considered a thin reconnection layer rather than a turbulence-broadened reconnection region. 
Accordingly, the acceleration with $r_g \gg \Delta$ happens outside the reconnection layer for the ``box" model.
Under the assumption that the accelerated particles can sample both the converging inflows and diverging outflows, 
the acceleration occurs only by virtue of fluid compression. With a constant density in an incompressible medium, 
the ``box" model leads to zero energy gain. 
As discussed in Section \ref{ssec:recacc},
the assumption of 
efficient scattering 
responsible for isotropizing the particle distribution
in the entire system that encompasses both inflows and outflows is invalid, as turbulence only exists within the reconnection region. 
In comparison, the turbulent reconnection acceleration with $r_g < \Delta$ considered in this work
happens within the reconnection region. 
Particles gain energy from the converging inflows,
irrespective of the fluid compressibility.}

\subsection{Comparison with other earlier studies on reconnection acceleration}

It is generally accepted that the acceleration by the motional electric field associated with reconnection-driven bulk flows
dominates over the acceleration by the non-ideal electric field at reconnection X-points
\citep{Guo19}.
As discussed in Section \ref{ssec:recacc},
in 2D reconnection models,
the curvature drift acceleration due to the motional electric field induced by contraction and merging of magnetic islands
is identified as the primary acceleration mechanism. 
{ In 3D reconnection, 
magnetic islands do not exist. 
The curvature drift acceleration 
associated with the motional electric field induced by outflows is only important when the reconnection is in the linear stage 
with insignificant turbulence and slow inflows
\citep{Bere16}.}
As the outflow speed only approaches $V_A$ at the reconnection exhausts, 
only a small fraction of particles in the outflows far away from the central reconnection region can gain energy from the kinetic energy of the outflows.
{ When the reconnection enters the nonlinear stage with significant turbulence and thus inflows, i.e., high reconnection rate, 
the magnetic energy release related to magnetic pressure gradients dominates over that related to magnetic curvature relaxation
\citep{Du22}.
The Fermi acceleration associated with the motional electric field induced by inflows becomes important, as indicated by
test particle simulations
\citep{Zjf22}.}

{ We do not consider the 2D reconnection with tearing instability and plasmoids, which are suppressed by other fluid instabilities, e.g., Kelvin-Helmholtz instability, 
and turbulence in 3D reconnection
\citep{Bere17,Kow20}.
The difference between 2D and 3D reconnection entails the difference between 2D and 3D reconnection acceleration.}
Unlike 2D reconnection where the reconnection sites, i.e., X-points, occupy a vanishing fraction of a large-scale system
\citep{Dah14},
3D turbulent reconnection has the volume-filling reconnection region, which drives significant inflows. Particles gain energy from the kinetic energy of the inflows 
\citep{DeG05},
and thus the acceleration is strongly dependent on the turbulent reconnection rate. 
The LV99 model we adopt is applicable to describing the fast reconnection events in solar flares 
\citep{Chi20}
and other high-energy astrophysical phenomena 
\citep{LazZ19}.
The corresponding reconnection rate is $\gtrsim 0.1$.
Under the consideration of fast reconnection, we expect that the acceleration 
arising from particles interacting with the inflows becomes the dominant acceleration mechanism. 
{ Its dominance over 
the curvature drift acceleration in 3D fast turbulent reconnection will be further numerically tested in our future work.}


In addition to the reconnection-driven turbulence, when there is an external source of turbulence, 
the reconnection rate can be further increased
(LV99; \citealt{KowL09}).
The externally driven turbulence can increase the range of inflow obliquities for 
more particles to undergo reflection at a small $\theta$.

Although the medium compressibility is not required for reconnection acceleration (see Section \ref{ssec:recacc}),
the increase of the turbulence driven by reconnection in a highly compressible medium 
(Kowal et al. in prep) 
is important for increasing the reconnection rate and thus the
acceleration efficiency.



\section{Conclusions}

Efficient reconnection acceleration of energetic particles
requires a global high reconnection rate on macroscopic scales. 
The macroscopic diffusion of turbulent magnetic fields is independent of microscopic magnetic diffusivities, and thus 
the 3D turbulent reconnection is a fast reconnection process. 
The macroscopic reconnection rate is high when the level of turbulence is high (LV99). 

Without invoking the compression of gas, 
particles that bounce between converging magnetic fields are accelerated by head-on collisions with the fast inflows, with the kinetic energy of the 
inflows converted to the particle energy. 
Furthermore, the reconnection-driven turbulence introduces a variety of local inflow obliquities. 
With the large inflow speed along the local magnetic field in the reconnection region, 
a reflected particle can occasionally gain significant energy after a single encounter with the inflow. 
The effect of fluid compressibility on acceleration is via its effect
on the reconnection rate.

Our study applies to the reconnection events with a high reconnection rate, 
where the acceleration arising from the interaction of particles with the inflows is important. 
We note that { in the linear stage of reconnection with a low 
reconnection rate,} the acceleration via the interaction with the outflows, i.e., 
the curvature drift acceleration, may become dominant 
\citep{Bere16}.

At a small $\theta$ with a strong guide field, the mirroring effect caused by magnetic reconnection
is weak, and the acceleration is dominated by that of unreflected particles. 
The unreflected particles are confined in the reconnection region due to the convergence of strong incoming magnetic fields.  
At a large $\theta$ with a weak guide field, 
the mirroring effect becomes strong,
allowing particles to be reflected at large inflow obliquities. 
The acceleration is dominated by that of reflected particles unless the inflow speed is a significant fraction of light speed.
In the latter case, the acceleration at a large $\theta$ is also dominated by that of unreflected particles. 
The confinement of reflected particles in the reconnection region is due to the mirror reflection. 
Both reflected and unreflected particles undergo the reconnection drift acceleration, which is caused by the gradient drift of particles along 
the motional electric field induced by the inflows.

Particles attached to turbulent magnetic fields can escape from the reconnection region by advection with the outflow.
The escape probability $P_\text{esp}$ turns out to depend on the inflow speed because of the mass conservation. 
Given the dependence of the averaged fractional energy gain $d$ and $P_\text{esp}$ on the inflow speed, 
the energy spectrum of accelerated particles does not have a universal spectral index. 
We find that it approximately falls in the range $2.5 \lesssim \zeta < 4$.
The steepest spectrum with $\zeta = 4$ is expected at a small $\theta$.

Scattering diffusion is not a necessary mechanism for confining particles in the reconnection region. 
Given a short crossing time of a particle through the reconnection region by ballistically following the turbulent magnetic field, 
the reconnection acceleration at a large $\theta$ and a large inflow speed can be very efficient, 
with the acceleration timescale comparable to the light crossing time of the thickness $\Delta$ of the reconnection region.
Moreover, the maximum particle energy is not limited by $\Delta$, as 
particles with $r_g > \Delta$ can also be accelerated 
\citep{oiLa12,Zhang21}. 
Given the macroscopic size of $\Delta$ of turbulent reconnection and strong magnetic fields in the inflows, 
particles can be accelerated to very high energies. 
For relativistic reconnection acceleration, particles are naturally trapped in the reconnection region by strong incoming magnetic fields 
and undergo successive acceleration with very significant energy gain. 
With all above factors considered, 
we see that 3D turbulent reconnection acceleration can be a very promising mechanism for producing ultra-high-energy cosmic rays
in strongly magnetized astrophysical plasmas.

\acknowledgments
S.X. acknowledges the support for 
this work provided by NASA through the NASA Hubble Fellowship grant \# HST-HF2-51473.001-A awarded by the Space Telescope Science Institute, which is operated by the Association of Universities for Research in Astronomy, Incorporated, under NASA contract NAS5-26555. 
A.L. acknowledges the support of NASA ATP  AAH7546.
\software{MATLAB \citep{MATLAB:2021}}

\bibliographystyle{aasjournal}
\bibliography{xu}

\begin{thebibliography}{}
\expandafter\ifx\csname natexlab\endcsname\relax\def\natexlab#1{#1}\fi
\providecommand{\url}[1]{\href{#1}{#1}}
\providecommand{\dodoi}[1]{doi:~\href{http://doi.org/#1}{\nolinkurl{#1}}}
\providecommand{\doeprint}[1]{\href{http://ascl.net/#1}{\nolinkurl{http://ascl.net/#1}}}
\providecommand{\doarXiv}[1]{\href{https://arxiv.org/abs/#1}{\nolinkurl{https://arxiv.org/abs/#1}}}

\bibitem[{{Armstrong} {et~al.}(1985){Armstrong}, {Pesses}, \& {Decker}}]{Arm85}
{Armstrong}, T.~P., {Pesses}, M.~E., \& {Decker}, R.~B. 1985, Washington DC
  American Geophysical Union Geophysical Monograph Series, 35, 271,
  \dodoi{10.1029/GM035p0271}

\bibitem[{{Axford} {et~al.}(1977){Axford}, {Leer}, \& {Skadron}}]{Axf77}
{Axford}, W.~I., {Leer}, E., \& {Skadron}, G. 1977, in International Cosmic Ray
  Conference, Vol.~11, International Cosmic Ray Conference, 132

\bibitem[{{Ball} \& {Melrose}(2001)}]{BaMel01}
{Ball}, L., \& {Melrose}, D.~B. 2001, \pasa, 18, 361, \dodoi{10.1071/AS01047}

\bibitem[{{Bell}(1978)}]{Bell78}
{Bell}, A.~R. 1978, \mnras, 182, 147, \dodoi{10.1093/mnras/182.2.147}

\bibitem[{{Beresnyak}(2017)}]{Bere17}
{Beresnyak}, A. 2017, \apj, 834, 47, \dodoi{10.3847/1538-4357/834/1/47}

\bibitem[{{Beresnyak} \& {Li}(2016)}]{Bere16}
{Beresnyak}, A., \& {Li}, H. 2016, \apj, 819, 90,
  \dodoi{10.3847/0004-637X/819/2/90}

\bibitem[{{Blandford} \& {Ostriker}(1978)}]{Blan78}
{Blandford}, R.~D., \& {Ostriker}, J.~P. 1978, \apjl, 221, L29,
  \dodoi{10.1086/182658}

\bibitem[{{Cassak} {et~al.}(2017){Cassak}, {Liu}, \& {Shay}}]{Cass17}
{Cassak}, P.~A., {Liu}, Y.~H., \& {Shay}, M.~A. 2017, Journal of Plasma
  Physics, 83, 715830501, \dodoi{10.1017/S0022377817000666}

\bibitem[{{Chitta} \& {Lazarian}(2020)}]{Chi20}
{Chitta}, L.~P., \& {Lazarian}, A. 2020, \apjl, 890, L2,
  \dodoi{10.3847/2041-8213/ab6f0a}

\bibitem[{Comisso \& Sironi(2018)}]{LucS18}
Comisso, L., \& Sironi, L. 2018, Phys. Rev. Lett., 121, 255101,
  \dodoi{10.1103/PhysRevLett.121.255101}

\bibitem[{{Dahlin} {et~al.}(2014){Dahlin}, {Drake}, \& {Swisdak}}]{Dah14}
{Dahlin}, J.~T., {Drake}, J.~F., \& {Swisdak}, M. 2014, Physics of Plasmas, 21,
  092304, \dodoi{10.1063/1.4894484}

\bibitem[{{Dahlin} {et~al.}(2017){Dahlin}, {Drake}, \& {Swisdak}}]{Dah17}
---. 2017, Physics of Plasmas, 24, 092110, \dodoi{10.1063/1.4986211}

\bibitem[{{de Gouveia dal Pino} \& {Lazarian}(2005)}]{DeG05}
{de Gouveia dal Pino}, E.~M., \& {Lazarian}, A. 2005, \aap, 441, 845,
  \dodoi{10.1051/0004-6361:20042590}

\bibitem[{{de Hoffmann} \& {Teller}(1950)}]{deTe50}
{de Hoffmann}, F., \& {Teller}, E. 1950, Physical Review, 80, 692,
  \dodoi{10.1103/PhysRev.80.692}

\bibitem[{{Decker}(1988)}]{Deck88}
{Decker}, R.~B. 1988, \ssr, 48, 195, \dodoi{10.1007/BF00226009}

\bibitem[{{Deng} {et~al.}(2016){Deng}, {Zhang}, {Zhang}, \& {Li}}]{Deng16}
{Deng}, W., {Zhang}, H., {Zhang}, B., \& {Li}, H. 2016, \apjl, 821, L12,
  \dodoi{10.3847/2041-8205/821/1/L12}

\bibitem[{{Drake} {et~al.}(2008){Drake}, {Shay}, \& {Swisdak}}]{Dra08}
{Drake}, J.~F., {Shay}, M.~A., \& {Swisdak}, M. 2008, Physics of Plasmas, 15,
  042306, \dodoi{10.1063/1.2901194}

\bibitem[{{Drake} {et~al.}(2006){Drake}, {Swisdak}, {Che}, \& {Shay}}]{Drak06}
{Drake}, J.~F., {Swisdak}, M., {Che}, H., \& {Shay}, M.~A. 2006, \nat, 443,
  553, \dodoi{10.1038/nature05116}

\bibitem[{{Drake} {et~al.}(2013){Drake}, {Swisdak}, \& {Fermo}}]{Drak13}
{Drake}, J.~F., {Swisdak}, M., \& {Fermo}, R. 2013, \apjl, 763, L5,
  \dodoi{10.1088/2041-8205/763/1/L5}

\bibitem[{{Drury}(1983)}]{Drury83}
{Drury}, L.~O. 1983, Reports on Progress in Physics, 46, 973,
  \dodoi{10.1088/0034-4885/46/8/002}

\bibitem[{{Drury}(2012)}]{Druf12}
---. 2012, \mnras, 422, 2474, \dodoi{10.1111/j.1365-2966.2012.20804.x}

\bibitem[{{Drury} {et~al.}(1999){Drury}, {Duffy}, {Eichler}, \&
  {Mastichiadis}}]{Drur99}
{Drury}, L.~O., {Duffy}, P., {Eichler}, D., \& {Mastichiadis}, A. 1999, \aap,
  347, 370.
\newblock \doarXiv{astro-ph/9905178}

\bibitem[{{Du} {et~al.}(2022){Du}, {Li}, {Fu}, {Gan}, \& {Li}}]{Du22}
{Du}, S., {Li}, H., {Fu}, X., {Gan}, Z., \& {Li}, S. 2022, \apj, 925, 128,
  \dodoi{10.3847/1538-4357/ac3de1}

\bibitem[{{Eyink} {et~al.}(2013){Eyink}, {Vishniac}, {Lalescu}, {Aluie},
  {Kanov}, {B{\"u}rger}, {Burns}, {Meneveau}, \& {Szalay}}]{Eyin13}
{Eyink}, G., {Vishniac}, E., {Lalescu}, C., {et~al.} 2013, \nat, 497, 466,
  \dodoi{10.1038/nature12128}

\bibitem[{{Eyink}(2015)}]{Eyink15}
{Eyink}, G.~L. 2015, \apj, 807, 137, \dodoi{10.1088/0004-637X/807/2/137}

\bibitem[{{Eyink} {et~al.}(2011){Eyink}, {Lazarian}, \& {Vishniac}}]{Eyink2011}
{Eyink}, G.~L., {Lazarian}, A., \& {Vishniac}, E.~T. 2011, \apj, 743, 51,
  \dodoi{10.1088/0004-637X/743/1/51}

\bibitem[{{Guo} {et~al.}(2014){Guo}, {Li}, {Daughton}, \& {Liu}}]{Guo14}
{Guo}, F., {Li}, H., {Daughton}, W., \& {Liu}, Y.-H. 2014, \prl, 113, 155005,
  \dodoi{10.1103/PhysRevLett.113.155005}

\bibitem[{{Guo} {et~al.}(2019){Guo}, {Li}, {Daughton}, {Kilian}, {Li}, {Liu},
  {Yan}, \& {Ma}}]{Guo19}
{Guo}, F., {Li}, X., {Daughton}, W., {et~al.} 2019, \apjl, 879, L23,
  \dodoi{10.3847/2041-8213/ab2a15}

\bibitem[{{Hu} {et~al.}(2022{\natexlab{a}}){Hu}, {Lazarian}, \& {Xu}}]{Hucr22}
{Hu}, Y., {Lazarian}, A., \& {Xu}, S. 2022{\natexlab{a}}, \mnras, 512, 2111,
  \dodoi{10.1093/mnras/stac319}

\bibitem[{{Hu} {et~al.}(2022{\natexlab{b}}){Hu}, {Xu}, {Stone}, \&
  {Lazarian}}]{Hu22sh}
{Hu}, Y., {Xu}, S., {Stone}, J.~M., \& {Lazarian}, A. 2022{\natexlab{b}},
  arXiv:2207.06941.
\newblock \doarXiv{2207.06941}

\bibitem[{{Jafari} {et~al.}(2021){Jafari}, {Vishniac}, \& {Xu}}]{Jaf21}
{Jafari}, A., {Vishniac}, E.~T., \& {Xu}, S. 2021, \apj, 906, 109,
  \dodoi{10.3847/1538-4357/abca47}

\bibitem[{{Jokipii}(1982)}]{Jok82}
{Jokipii}, J.~R. 1982, \apj, 255, 716, \dodoi{10.1086/159870}

\bibitem[{{Jokipii}(1987)}]{Jok87}
---. 1987, \apj, 313, 842, \dodoi{10.1086/165022}

\bibitem[{{Jones} \& {Ellison}(1991)}]{Jon91}
{Jones}, F.~C., \& {Ellison}, D.~C. 1991, \ssr, 58, 259,
  \dodoi{10.1007/BF01206003}

\bibitem[{{Kadowaki} {et~al.}(2021){Kadowaki}, {de Gouveia Dal Pino},
  {Medina-Torrej{\'o}n}, {Mizuno}, \& {Kushwaha}}]{Kado21}
{Kadowaki}, L. H.~S., {de Gouveia Dal Pino}, E.~M., {Medina-Torrej{\'o}n},
  T.~E., {Mizuno}, Y., \& {Kushwaha}, P. 2021, \apj, 912, 109,
  \dodoi{10.3847/1538-4357/abee7a}

\bibitem[{{Kirk} \& {Heavens}(1989)}]{Kirk89}
{Kirk}, J.~G., \& {Heavens}, A.~F. 1989, \mnras, 239, 995,
  \dodoi{10.1093/mnras/239.3.995}

\bibitem[{{Kowal} {et~al.}(2011){Kowal}, {de Gouveia Dal Pino}, \&
  {Lazarian}}]{Kow11}
{Kowal}, G., {de Gouveia Dal Pino}, E.~M., \& {Lazarian}, A. 2011, \apj, 735,
  102, \dodoi{10.1088/0004-637X/735/2/102}

\bibitem[{{Kowal} {et~al.}(2012{\natexlab{a}}){Kowal}, {de Gouveia Dal Pino},
  \& {Lazarian}}]{Kow12}
---. 2012{\natexlab{a}}, Physical Review Letters, 108, 241102,
  \dodoi{10.1103/PhysRevLett.108.241102}

\bibitem[{{Kowal} {et~al.}(2017){Kowal}, {Falceta-Gon{\c c}alves}, {Lazarian},
  \& {Vishniac}}]{Kow17}
{Kowal}, G., {Falceta-Gon{\c c}alves}, D.~A., {Lazarian}, A., \& {Vishniac},
  E.~T. 2017, \apj, 838, 91, \dodoi{10.3847/1538-4357/aa6001}

\bibitem[{{Kowal} {et~al.}(2020){Kowal}, {Falceta-Gon{\c{c}}alves}, {Lazarian},
  \& {Vishniac}}]{Kow20}
{Kowal}, G., {Falceta-Gon{\c{c}}alves}, D.~A., {Lazarian}, A., \& {Vishniac},
  E.~T. 2020, \apj, 892, 50, \dodoi{10.3847/1538-4357/ab7a13}

\bibitem[{{Kowal} {et~al.}(2009){Kowal}, {Lazarian}, {Vishniac}, \&
  {Otmianowska-Mazur}}]{KowL09}
{Kowal}, G., {Lazarian}, A., {Vishniac}, E.~T., \& {Otmianowska-Mazur}, K.
  2009, in Revista Mexicana de Astronomia y Astrofisica Conference Series,
  Vol.~36, Revista Mexicana de Astronomia y Astrofisica Conference Series,
  89--96.
\newblock \doarXiv{0812.2024}

\bibitem[{{Kowal} {et~al.}(2012{\natexlab{b}}){Kowal}, {Lazarian}, {Vishniac},
  \& {Otmianowska-Mazur}}]{KL12}
{Kowal}, G., {Lazarian}, A., {Vishniac}, E.~T., \& {Otmianowska-Mazur}, K.
  2012{\natexlab{b}}, Nonlinear Processes in Geophysics, 19, 297,
  \dodoi{10.5194/npg-19-297-2012}

\bibitem[{{Krauss-Varban} \& {Wu}(1989)}]{Krauss89}
{Krauss-Varban}, D., \& {Wu}, C.~S. 1989, \jgr, 94, 15367,
  \dodoi{10.1029/JA094iA11p15367}

\bibitem[{{Krymsky}(1977)}]{Kry77}
{Krymsky}, G.~F. 1977, Dokl. Akad. Nauk SSSR, 234, 1306

\bibitem[{{Lazarian}(2005)}]{Laz05}
{Lazarian}, A. 2005, in American Institute of Physics Conference Series, Vol.
  784, Magnetic Fields in the Universe: From Laboratory and Stars to Primordial
  Structures., ed. E.~M. {de Gouveia dal Pino}, G.~{Lugones}, \& A.~{Lazarian},
  42--53, \dodoi{10.1063/1.2077170}

\bibitem[{{Lazarian}(2014)}]{Laz14r}
{Lazarian}, A. 2014, \ssr, 181, 1, \dodoi{10.1007/s11214-013-0031-5}

\bibitem[{{Lazarian} {et~al.}(2020){Lazarian}, {Eyink}, {Jafari}, {Kowal},
  {Li}, {Xu}, \& {Vishniac}}]{Laz20}
{Lazarian}, A., {Eyink}, G.~L., {Jafari}, A., {et~al.} 2020, Physics of
  Plasmas, 27, 012305, \dodoi{10.1063/1.5110603}

\bibitem[{{Lazarian} {et~al.}(2012){Lazarian}, {Kowal}, {de Gouveia Dal Pino},
  \& {Vishniac}}]{oiLa12}
{Lazarian}, A., {Kowal}, G., {de Gouveia Dal Pino}, E., \& {Vishniac}, E.~T.
  2012, in Astrophysics and Space Science Proceedings, Vol.~33, Multi-scale
  Dynamical Processes in Space and Astrophysical Plasmas, 11,
  \dodoi{10.1007/978-3-642-30442-2\_2}

\bibitem[{{Lazarian} \& {Opher}(2009)}]{LaO09}
{Lazarian}, A., \& {Opher}, M. 2009, \apj, 703, 8,
  \dodoi{10.1088/0004-637X/703/1/8}

\bibitem[{{Lazarian} \& {Vishniac}(2000)}]{LV00}
{Lazarian}, A., \& {Vishniac}, E. 2000, in Revista Mexicana de Astronomia y
  Astrofisica Conference Series, Vol.~9, Revista Mexicana de Astronomia y
  Astrofisica Conference Series, ed. S.~J. {Arthur}, N.~S. {Brickhouse}, \&
  J.~{Franco}, 55--62

\bibitem[{{Lazarian} \& {Vishniac}(1999)}]{LV99}
{Lazarian}, A., \& {Vishniac}, E.~T. 1999, \apj, 517, 700,
  \dodoi{10.1086/307233}

\bibitem[{{Lazarian} {et~al.}(2004){Lazarian}, {Vishniac}, \& {Cho}}]{LVC04}
{Lazarian}, A., {Vishniac}, E.~T., \& {Cho}, J. 2004, \apj, 603, 180,
  \dodoi{10.1086/381383}

\bibitem[{{Lazarian} \& {Xu}(2021{\natexlab{a}})}]{LX21m}
{Lazarian}, A., \& {Xu}, S. 2021{\natexlab{a}}, \apj, 923, 53,
  \dodoi{10.3847/1538-4357/ac2de9}

\bibitem[{{Lazarian} \& {Xu}(2021{\natexlab{b}})}]{LX21}
---. 2021{\natexlab{b}}, submitted to ApJ

\bibitem[{{Lazarian} {et~al.}(2019){Lazarian}, {Zhang}, \& {Xu}}]{LazZ19}
{Lazarian}, A., {Zhang}, B., \& {Xu}, S. 2019, \apj, 882, 184,
  \dodoi{10.3847/1538-4357/ab2b38}

\bibitem[{{le Roux} {et~al.}(2018){le Roux}, {Zank}, \& {Khabarova}}]{leRZ18}
{le Roux}, J.~A., {Zank}, G.~P., \& {Khabarova}, O.~V. 2018, \apj, 864, 158,
  \dodoi{10.3847/1538-4357/aad8b3}

\bibitem[{{Lee} \& {Fisk}(1982)}]{Lee82}
{Lee}, M.~A., \& {Fisk}, L.~A. 1982, \ssr, 32, 205, \dodoi{10.1007/BF00225185}

\bibitem[{{Li} {et~al.}(2021){Li}, {Guo}, \& {Liu}}]{Li21}
{Li}, X., {Guo}, F., \& {Liu}, Y.-H. 2021, Physics of Plasmas, 28, 052905,
  \dodoi{10.1063/5.0047644}

\bibitem[{Longair(1997)}]{Longairbook}
Longair, M.~S. 1997, High energy astrophysics. Volume 2: Stars, the galaxy and
  the interstellar medium (Cambridge U Press, 1994)

\bibitem[{{Lyutikov} {et~al.}(2018){Lyutikov}, {Komissarov}, {Sironi}, \&
  {Porth}}]{Lyut18}
{Lyutikov}, M., {Komissarov}, S., {Sironi}, L., \& {Porth}, O. 2018, Journal of
  Plasma Physics, 84, 635840201, \dodoi{10.1017/S0022377818000168}

\bibitem[{{Mandt} {et~al.}(1994){Mandt}, {Denton}, \& {Drake}}]{Man94}
{Mandt}, M.~E., {Denton}, R.~E., \& {Drake}, J.~F. 1994, \grl, 21, 73,
  \dodoi{10.1029/93GL03382}

\bibitem[{{Marcowith} {et~al.}(2020){Marcowith}, {Ferrand}, {Grech}, {Meliani},
  {Plotnikov}, \& {Walder}}]{Mar20}
{Marcowith}, A., {Ferrand}, G., {Grech}, M., {et~al.} 2020, Living Reviews in
  Computational Astrophysics, 6, 1, \dodoi{10.1007/s41115-020-0007-6}

\bibitem[{MATLAB(2021)}]{MATLAB:2021}
MATLAB. 2021, MATLAB and Statistics Toolbox Release 2021b (Natick,
  Massachusetts: The MathWorks Inc.)

\bibitem[{{Naito} \& {Takahara}(1995)}]{NaTa95}
{Naito}, T., \& {Takahara}, F. 1995, Progress of Theoretical Physics, 93, 287,
  \dodoi{10.1143/PTP.93.287}

\bibitem[{{Ostrowski}(1988)}]{Ostr88}
{Ostrowski}, M. 1988, \mnras, 233, 257, \dodoi{10.1093/mnras/233.2.257}

\bibitem[{{Parker}(1957)}]{Pars57}
{Parker}, E.~N. 1957, \jgr, 62, 509, \dodoi{10.1029/JZ062i004p00509}

\bibitem[{{Parker}(1993)}]{Par93}
---. 1993, \apj, 414, 389, \dodoi{10.1086/173085}

\bibitem[{{Parnell} {et~al.}(2011){Parnell}, {Maclean}, {Haynes}, \&
  {Galsgaard}}]{Parn11}
{Parnell}, C.~E., {Maclean}, R.~C., {Haynes}, A.~L., \& {Galsgaard}, K. 2011,
  in Astrophysical Dynamics: From Stars to Galaxies, ed. N.~H. {Brummell},
  A.~S. {Brun}, M.~S. {Miesch}, \& Y.~{Ponty}, Vol. 271, 227--238,
  \dodoi{10.1017/S1743921311017650}

\bibitem[{{Priest} {et~al.}(2003){Priest}, {Hornig}, \& {Pontin}}]{Prie03}
{Priest}, E.~R., {Hornig}, G., \& {Pontin}, D.~I. 2003, Journal of Geophysical
  Research (Space Physics), 108, 1285, \dodoi{10.1029/2002JA009812}

\bibitem[{{Santos-Lima} {et~al.}(2010){Santos-Lima}, {Lazarian}, {de Gouveia
  Dal Pino}, \& {Cho}}]{Sant10}
{Santos-Lima}, R., {Lazarian}, A., {de Gouveia Dal Pino}, E.~M., \& {Cho}, J.
  2010, \apj, 714, 442, \dodoi{10.1088/0004-637X/714/1/442}

\bibitem[{{Shibata}(2020)}]{Shib20}
{Shibata}, K. 2020, in AAS/Solar Physics Division Meeting, Vol.~52, AAS/Solar
  Physics Division Meeting, 100.01

\bibitem[{{Sironi} \& {Spitkovsky}(2014)}]{Sir14}
{Sironi}, L., \& {Spitkovsky}, A. 2014, \apjl, 783, L21,
  \dodoi{10.1088/2041-8205/783/1/L21}

\bibitem[{{Sonnerup}(1969)}]{Son69}
{Sonnerup}, B.~U.~{\"O}. 1969, \jgr, 74, 1301, \dodoi{10.1029/JA074i005p01301}

\bibitem[{{Spiegel}(1971)}]{Spie71}
{Spiegel}, E.~A. 1971, \araa, 9, 323,
  \dodoi{10.1146/annurev.aa.09.090171.001543}

\bibitem[{{Sweet}(1958)}]{Swe58}
{Sweet}, P.~A. 1958, The Observatory, 78, 30

\bibitem[{{Takamoto} {et~al.}(2015){Takamoto}, {Inoue}, \& {Lazarian}}]{Tak15}
{Takamoto}, M., {Inoue}, T., \& {Lazarian}, A. 2015, \apj, 815, 16,
  \dodoi{10.1088/0004-637X/815/1/16}

\bibitem[{{Turkmani} {et~al.}(2005){Turkmani}, {Vlahos}, {Galsgaard},
  {Cargill}, \& {Isliker}}]{TUrk05}
{Turkmani}, R., {Vlahos}, L., {Galsgaard}, K., {Cargill}, P.~J., \& {Isliker},
  H. 2005, \apjl, 620, L59, \dodoi{10.1086/428395}

\bibitem[{{Uritsky} {et~al.}(2017){Uritsky}, {Roberts}, {DeVore}, \&
  {Karpen}}]{Uri17}
{Uritsky}, V.~M., {Roberts}, M.~A., {DeVore}, C.~R., \& {Karpen}, J.~T. 2017,
  \apj, 837, 123, \dodoi{10.3847/1538-4357/aa5cb9}

\bibitem[{{Werner} \& {Uzdensky}(2021)}]{Wern21}
{Werner}, G.~R., \& {Uzdensky}, D.~A. 2021, Journal of Plasma Physics, 87,
  905870613, \dodoi{10.1017/S0022377821001185}

\bibitem[{{Wu}(1984)}]{Wu84}
{Wu}, C.~S. 1984, \jgr, 89, 8857, \dodoi{10.1029/JA089iA10p08857}

\bibitem[{{Xu}(2022)}]{Xu22}
{Xu}, S. 2022, \apj, 934, 136, \dodoi{10.3847/1538-4357/ac7c68}

\bibitem[{{Xu} \& {Lazarian}(2016)}]{XL16}
{Xu}, S., \& {Lazarian}, A. 2016, \apj, 833, 215,
  \dodoi{10.3847/1538-4357/833/2/215}

\bibitem[{{Xu} \& {Lazarian}(2017)}]{XuL17}
---. 2017, \apj, 850, 126, \dodoi{10.3847/1538-4357/aa956b}

\bibitem[{{Xu} \& {Lazarian}(2022)}]{XL22}
---. 2022, \apj, 925, 48, \dodoi{10.3847/1538-4357/ac3824}

\bibitem[{{Xu} \& {Yan}(2013)}]{XY13}
{Xu}, S., \& {Yan}, H. 2013, \apj, 779, 140,
  \dodoi{10.1088/0004-637X/779/2/140}

\bibitem[{{Zhang} \& {Yan}(2011)}]{Zh11}
{Zhang}, B., \& {Yan}, H. 2011, \apj, 726, 90,
  \dodoi{10.1088/0004-637X/726/2/90}

\bibitem[{{Zhang} {et~al.}(2021{\natexlab{a}}){Zhang}, {Sironi}, \&
  {Giannios}}]{Zhang21}
{Zhang}, H., {Sironi}, L., \& {Giannios}, D. 2021{\natexlab{a}}, \apj, 922,
  261, \dodoi{10.3847/1538-4357/ac2e08}

\bibitem[{{Zhang} {et~al.}(2022){Zhang}, {Kowal}, {Lazarian}, \& {Xu}}]{Zjf22}
{Zhang}, J., {Kowal}, G., {Lazarian}, A., \& {Xu}, S. 2022, submitted to MNRAS

\bibitem[{{Zhang} {et~al.}(2021{\natexlab{b}}){Zhang}, {Guo}, {Daughton}, {Li},
  \& {Li}}]{Zhg21}
{Zhang}, Q., {Guo}, F., {Daughton}, W., {Li}, H., \& {Li}, X.
  2021{\natexlab{b}}, \prl, 127, 185101, \dodoi{10.1103/PhysRevLett.127.185101}

\end{thebibliography}

\end{document}